\documentclass{tmsce}

\usepackage[utf8]{inputenc}
\usepackage{lipsum}
\usepackage{amsmath,amsfonts,amsthm}
\usepackage{algorithmic}
\usepackage{algorithm}
\usepackage{array}
\usepackage{textcomp}
\usepackage{stfloats}
\usepackage{url}
\usepackage{verbatim}
\usepackage{graphicx}
\usepackage{cite}
\usepackage{mwe} 

\usepackage{amssymb}

\usepackage{physics}
\usepackage{mathtools}
\usepackage{tikz, tikz-qtree}
\usepackage{graphicx}
\usepackage{subcaption}
\usetikzlibrary{calc}

\numberwithin{equation}{section}
\makeatletter
\renewcommand{\tmsce@topstrip}{}
\makeatother
\title{Generalized Numerical Construction of MUBs: A Group Theoretical Investigation}

\authors{B. Gültekin$^{1}$, S. B. Samuel$^{2}$ and Z. Gedik$^{1}$}
\affiliation{$^1$Faculty of Engineering and Natural Sciences, Sabanci University, Istanbul, Turkey \\
$^2$ Department of Theoretical Physics, Budapest University of Technology and Economics, Budapest, Hungary \\
\textit{Corresponding author:} bugra.gultekin@sabanciuniv.edu}

\begin{document}

\maketitle

\begin{abstract}

Mutually Unbiased Bases (MUBs) constitute a fundamental geometric structure in quantum theory, known for providing an optimal measurement scheme for quantum state tomography. In prime and prime-power dimensions, analytical constructions of maximal sets of MUBs are well-known and standard construction relies on the Weyl-Heisenberg (WH) group and finite fields. In non-prime-power dimensions, on the other hand, the existence of such maximal sets remains an open question. We present a generalized numerical method of constructing MUBs without any reliance on \textit{a priori} group structure or specific algebraic frameworks. Formulating the problem at the level of Gram matrix, we reduce the search for complete sets of d+1 MUBs to a phase space optimisation problem. We use the fact that the MUB Gram matrix is a projection matrix, and the third- and fourth-order trace constraints are necessary and sufficient conditions for a projection matrix. We further develop a classification framework based on third-order Bargmann invariants and automorphism groups, allowing us to probe the underlying algebraic and geometric structure of the resulting configurations. Numerical applications of this method in dimensions $3$, $4$, and $5$ demonstrate that all numerically constructed solutions are mutually isomorphic, are isolated points in phase space, and possess automorphism groups that coincide exactly with the Clifford group, the normalizer of the WH group. In the lowest non-prime-power dimension $6$ where the existence of a complete set remains open, our numerical search yields no solution. However, for incomplete sets of solutions, our search finds no set of four MUBs, but it yields several inequivalent triples, which we classify and whose neighbourhood we then investigate.

\end{abstract}

\tmsceendfrontmatter

\section{Introduction} \label{section:intro}


Quantum formalism admits a variety of geometric structures in the description of physical systems and measurements for obtaining the probability distribution of particular experimental outcomes. Among these geometric structures that hold significant importance in quantum foundations, quantum information, and mathematics are \textit{Mutually Unbiased Bases} (MUBs).

Consider a two-dimensional quantum system, such as a spin-$1/2$ particle, or any qubit system. A projective measurement along the $z$-direction corresponds to the basis $\mathcal{B}_z = \{\ket{0}, \ket{1}\}$, whose elements represent spin up and down along $z$. If the system is prepared in a state $\ket{\psi}$, the probabilities of obtaining these outcomes are given by Born's rule as $|\braket{0}{\psi}|^2$ and $|\braket{1}{\psi}|^2$. One may instead do a measurement along the $x$- or $y$-directions, corresponding to the bases $\mathcal{B}_x = \{\ket{+}, \ket{-}\}$ and $\mathcal{B}_y = \{\ket{i}, \ket{-i}\}$. If a measurement in the $z$-basis yields the outcome $\ket{0}$, i.e., if the system is prepared in the state $\ket{0}$, then subsequent measurements in either the $x$- or $y$-basis produce completely random outcomes, with $|\braket{+}{0}|^2 = |\braket{-}{0}|^2 = 1/2$, and similarly for $\mathcal{B}_y$. Thus, precise knowledge of the state with respect to one basis enforces \textit{maximal uncertainty} with respect to the others. This property is a finite-dimensional manifestation of Bohr's \textit{complementarity} — akin to position-momentum relation $|\braket{x}{p}|^2 = 1/2\pi$ \cite{Bengtsson2007,Bengtsson2006}. These bases — $\mathcal{B}_z, \mathcal{B}_x, \mathcal{B}_y$ — are mutually unbiased or complementary to one another.

In fact, in a $d$-dimensional Hilbert space, $\mathcal{H}^d$, any two bases $\mathcal{B}_\mu=\{\ket{\psi_{i}^\mu}\}$ and $\mathcal{B}_\nu=\{\ket{\psi_{j}^\nu}\}$ are mutually unbiased if the basis elements are related as
\begin{equation} \label{intro:MUBs}
\begin{aligned}
|\langle \psi_i^\mu | \psi_j^\nu \rangle|^2 = \begin{cases} \delta_{ij}, & \mu=\nu; \\ \frac{1}{d}, & \mu \neq \nu, \end{cases}
\end{aligned}
\end{equation}
where $\delta_{ij}$ is the Kronecker delta function and $1 \leq i,j \leq d$. 
Although the term was first coined by \cite{Wooters1989}, its conception dates back to Schwinger's work on unitary operator bases \cite{Schwinger1960} in which he demonstrated that one can construct particular observable operators in finite-dimensional Hilbert spaces whose eigenbases exhibit the property that measurement outcomes in one basis are uniformly distributed when the system is prepared in an eigenstate of the other — precisely the defining property of MUBs. Later, in the 1980s, Ivanović's pioneering work \cite{Ivonovic1981} discovered the importance of such bases in optimal \textit{quantum state tomography}. The description of a $d$-dimensional quantum system depends on $d^2-1$ independent parameters, as the density operator, with which a quantum system is described, is Hermitian and unit-trace. Measurement in a single orthonormal basis yields $d-1$ parameters, since those $d$ elements are constrained by completeness condition. Consequently, the minimum number of bases required to fully describe a $d$-dimensional system is $(d^2 - 1)/(d - 1) = d + 1$. Ivanović, and subsequently Wootters and Fields, showed that a set of $d+1$ MUBs achieves this bound optimally, providing a maximally efficient measurement scheme for complete state determination. This is what makes the number $d+1$ so significant in the study of MUBs.

This brings forth the main question that drives the research on the MUBs within the quantum information and mathematics community: Does the maximal number $d+1$ of MUBs exist in every dimension $d$? In prime and prime-power dimensions, we have recipes with which we can construct maximal number of MUBs analytically. We briefly overview one such recipe in Section \ref{section:knownSols} — operator algebraic method based on the WH group.

In non-prime-power dimensions, however, no general analytical method for constructing a maximal set of $d+1$ MUBs is known \cite{Bengtsson2007,Bengtsson2006,Durt2010}. In the lowest non-prime-power dimension, $d=6$, several numerical investigations suggest that at most $3$ MUBs can be found \cite{Butterley2007,Brierley2008,Grassl2009,Raynal2011}. It is widely suspected that the maximal number $d+1=7$ does not exist in dimension $6$, although there is no established proof of this claim \cite{Bengtsson2007,Bengtsson2006,Durt2010}. See \cite{McNulty2026s} for a comprehensive review of the MUB problem in non-prime-power dimensions.

A closely related problem concerns the existence of highly symmetric generalized measurements known as \textit{symmetric informationally complete positive-operator-valued measures} (SIC-POVMs), which correspond geometrically to $d^2$ equiangular lines in $d$-dimensional complex space \cite{ScottGrassl2010}. In his 1999 doctoral thesis, Zauner conjectured the existence of such structures in all finite dimensions. Despite substantial numerical evidence in moderately high dimensions, no general analytic proof of Zauner's conjecture is known \cite{Fuchs2017}.

MUBs and SIC-POVMs share several structural similarities. In geometric terms, in the case of MUBs, one constructs collections of orthonormal bases whose vectors are equiangular across distinct bases, whereas a SIC-POVM is a distribution of a single set of vectors as uniformly as possible in the complex space, achieving a single set of equiangular vectors. They both constitute a mathematical structure called a \textit{tight frame} \cite{Waldron2018}. The operator-algebraic construction of both structures, in known analytical methods, rely on WH group: MUBs are eigenstates of the commuting subgroups of WH operators, whereas SIC-POVMs are WH covariant orbit of a single fiducial state \cite{Bengtsson2017}. Moreover, they are linked through geometric structures known as \textit{finite affine planes} \cite{Grassl2009, Wootters2006}. 

In a recent work by Samuel and Gedik \cite{samuel2024}, a general framework was developed for classifying SIC-POVMs beyond standard covariance assumptions. That approach introduced a construction method independent of WH covariance and analyzed the associated symmetry groups, allowing for a novel way of classifying SIC-POVMs. 

In this work, we extend this framework to the problem of MUBs. Without any dependence on WH group or finite fields, we develop a Gram-matrix-based approach leading to what we term \textit{generalized numerical MUBs}. Because the construction presupposes no group algebraic structure, the framework can be applied in all dimensions — including the non-prime-power case, where every known analytic construction fails precisely because the structures it relies on are absent — and, through the gauge-invariant group theoretical classification, certifies whether any configuration found is equivalent to a known solution or constitutes a genuinely new structure. Rather than as a practical method of constructing MUBs in an arbitrary dimension, this generalized method is developed for investigating their geometric and group-theoretic structure under the fewest possible assumptions, capable of probing configurations that lie beyond the structural limitations built into the existing approaches.

We will show that the properties of a valid MUB Gram matrix would leave us with two constraints — third and fourth order traces of the MUB Gram matrix — which is sufficient to determine the geometry corresponding to a MUB structure (Section \ref{section:construction}). These solutions are characterized through \textit{third-order Bargmann invariants}, i.e., \textit{triple products}, and their \textit{automorphism groups}, allowing us to probe the underlying algebraic and geometric structure of the resulting configurations (Section \ref{section:classification}). 

In practice, inequivalent forms of MUBs carry differences not merely mathematical, but physical. Distinct MUB structures have different information extracting capabilities \cite{Yan2024}. In other words, such inequivalent complete sets amounts to genuinely different statistical performance. A method capable of constructing and classifying MUBs without presupposing a particular algebraic structure, is therefore of interest beyond the purely mathematical. Moreover, the Bargmann invariants underlying our classification have recently been shown to be experimentally measurable \cite{Oszmaniec2024, Wagner2024}, through which our classification may find relevance wherever MUBs are employed.

Furthermore, in addition to the structural similarities between SIC-POVMs and MUBs noted above, several works have sought to connect the two structures more explicitly. These take various forms: the common framework of WH group frames \cite{Feng2025}, the expansion of MUB projectors in terms of SIC-POVM elements \cite{Canturk2023}, and their experimental comparison through their application for quantum tomography \cite{Bent2015}. Whether such links may extend to a direct relation at the level of the Gram matrix and, in particular, whether the MUB existence problem can be recast in terms of a known SIC-POVM structure remains an open problem for further scrutiny.

Since the construction presupposes no group-algebraic structure, it explores the full solution space of MUB configurations and is, in principle, capable of revealing configurations other than the known solutions. In dimensions $3$, $4$, and $5$ we find that it recovers exclusively the known solutions, with automorphism groups coinciding with the Clifford group (Section \ref{section:results}). As we argue below, this result — obtained without any \textit{a priori} structural assumption — is consistent with the known uniqueness of complete MUB sets in dimensions $d \leq 5$ \cite{Blanchfield2014, Brierley2009}. It further confirms that the automorphism group of these configurations coincides with the Clifford group \cite{ChienWaldron2018}.

In dimension 6, the lowest non-prime-power dimension, our search for a complete set of $d+1 = 7$ MUBs yields no solution, though it is limited in scope by the substantially longer computation time. The method also applies to incomplete sets: the search for $3$ MUBs produce solutions with notable structural variety, whereas the search for $4$ MUBs, whose existence remains an open problem, yields none, reinforcing the long-standing suspicion that no more than $3$ MUBs exist in dimension $6$. Moreover, we characterize the triples of MUBs found by the generalized numerical search as well as some solutions in the literature to contrast them with what is found. Lastly, we investigate the neighbourhood of the solutions using the first-order perturbation method described in Section \ref{section:vicinity}.

\section{Analytical Construction} \label{section:knownSols}

The known methods for constructing standard MUBs span a wide range of mathematical frameworks including operator-algebraic approach based on the WH group \cite{Bandyopadhyay2002, Schwinger1960}; combinatorial designs involving relative difference sets, finite semifields, and symplectic spreads \cite{Godsil2009, Calderbank1997}; finite geometry, mutually orthogonal Latin squares, affine and projective planes \cite{Bengtsson2006, Bengtsson2017}; Galois rings \cite{Klappnecker2003}. Here, we briefly review the construction introduced by Bandyopadhyay et al. \cite{Bandyopadhyay2002}.


\subsection{Operator Algebraic Method} \label{section:knownSolsOperator}

Consider an orthogonal basis of unitary operators spanning the $d^2$-dimensional linear operator space $\mathbb{M}^d(\mathbb{C})$,
\begin{equation} 
\begin{aligned} 
B_{\mathbb{M}^d} = \{ U_1, U_2, \dots, U_{d^2} \},
\end{aligned}
\end{equation}
where we assume $U_1 = \mathbb{I}_d$, the identity operator of order $d$, and orthogonality is with respect to the Hilbert-Schmidt inner product. We partition this basis into $d+1$ disjoint classes, denoted by $\mathcal{C}$, as
\begin{equation} \label{partitionClasses}
\begin{aligned} 
B_{\mathbb{M}^d} = \{\mathbb{I}_d\} \cup \mathcal{C}_1 \cup \mathcal{C}_2 \cup \cdots \cup \mathcal{C}_{d+1},
\end{aligned}
\end{equation}
such that each class
\begin{equation}
\begin{aligned} 
\mathcal{C}_{j} = \{ U_{1}^{j}, \cdots,  U_{t}^{j}, \cdots, U_{d-1}^{j}\}
\end{aligned}
\end{equation}
consists of $d-1$ pairwise commuting operators: $\left[ U_{k}^{j},U_{k'}^{j} \right] = 0$. It was proven in \cite{Bandyopadhyay2002} that if such a partition exists, the common eigenbases associated with distinct classes are mutually unbiased. Such a basis is called a \textit{maximal commuting orthogonal basis}. The WH operators precisely satisfy the orthogonality and commutativity conditions established here.

\subsubsection{\textbf{\textit{Weyl-Heisenberg group}}}  In any dimension $d$, the generalized Pauli operators are defined as
\begin{equation}  \label{generalizedPauliOperators}
\begin{aligned}
X = \sum_{k=0}^{d-1} \ket{k+1} \bra{k}, \ \ \ \ Z = \sum_{k=0}^{d-1} e^{\frac{i2\pi k}{d}} \ket{k} \bra{k},
\end{aligned}
\end{equation}
where the index $k$ is taken as modulo $d$. The operators $X$ and $Z$, referred to as the \textit{shift} and \textit{phase} operators respectively, generate the WH group and constitute a natural extension of the two-dimensional Pauli operators — hence the name. We define the \textit{displacement operators} as
\begin{equation}
\begin{aligned}
D_{\textbf{p}} = \tau^{p_1 p_2} X^{p_1} Z^{p_2}, \qquad \textbf{p}=(p_1,p_2) \in \mathbb{Z}_d^2,
\end{aligned}
\end{equation}
where $\tau = -e^{i \pi / d}$ is a fixed phase, and $\mathbb{Z}_d$ denotes integers modulo $d$. WH group, $\textbf{H}(d)$, is formed by the displacement operators \cite{ScottGrassl2010}:
\begin{equation} \label{WHgroupDefinition}
\begin{aligned}
\textbf{H}(d) = \{ e^{i k} D_{\textbf{p}} : k \in \mathbb{Z}_d, \textbf{p} \in \mathbb{Z}_d^2 \}.
\end{aligned}
\end{equation}

\subsubsection{\textbf{Prime dimensions}} In prime number dimensions, the maximal commuting orthogonal basis is of the WH form:
\begin{equation} \label{operatorsPrimeDim}
B_{\mathbb{M}^d} = \{ X^{p_1}Z^{p_2} \}_{p_1,p_2=0}^{d-1}.
\end{equation}
These operators can be partitioned into maximal commuting classes by selecting pairs satisfying $\left[ D_{\textbf{p}}, D_{\textbf{q}} \right] = 0$ — yielding the constraint $p_1 q_2 - p_2 q_1 \equiv 0 \ (\text{mod} \ d)$. In dimension $d=3$, for example, the maximal commuting basis
\begin{equation}
\begin{aligned}
\mathcal{B}_{\mathbb{M}^3}= \{ \mathbb{I}_3,X,Z,X^2,Z^2,XZ,X^2Z,XZ^2,X^2Z^2 \},
\end{aligned}
\end{equation}
can be partitioned into commuting classes
\begin{equation} \label{class3D}
\begin{aligned}
\mathcal{C}_{1} &= \{ Z, Z^2 \} \\
\mathcal{C}_{2} &= \{ X, X^2 \} \\
\mathcal{C}_{3} &= \{ XZ, X^2Z^2 \} \\
\mathcal{C}_{4} &= \{ XZ^2, X^2Z \}.
\end{aligned}
\end{equation}

\begin{figure}[h] 
\centering
\begin{tikzpicture}[scale=1.25, every node/.style={font=\small}]
  \foreach \x in {0,1,2} \foreach \y in {0,1,2}{
    \filldraw[white, draw=black, line width=0.6pt] (\x,\y) circle (2.3pt);
  }
  \fill (0,0) circle (2.6pt);

  \foreach \y in {0,1,2}{
    \draw[line width=1.0pt] (0,\y) circle (2.3pt);
  }
  \foreach \x in {0,1,2}{
    \draw[line width=1.0pt] (\x,0) circle (2.3pt);
  }
  \foreach \t in {0,1,2}{
    \draw[line width=1.0pt] (\t,\t) circle (2.3pt);
  }
  \foreach \t/\u in {0/0,1/2,2/1}{
    \draw[dashed, line width=1.0pt] (\t,\u) circle (2.3pt);
  }

  \draw[line width=1.1pt] (0,0)--(0,2) node[above] {$\mathcal C_{1}$};
  \draw[line width=1.1pt] (0,0)--(2,0) node[right] {$\mathcal C_{2}$};
  \draw[line width=1.1pt] (0,0)--(2,2) node[above right] {$\mathcal C_{3}$};
  \draw[dashed, line width=1.1pt] (0,0)--(1,2) node[above right] {$\mathcal C_{4}$};
  \draw[dashed, line width=1.1pt] (1.5,0)--(2,1) node[above right] {$\mathcal C_{4}$};

\node[below left=-1pt] at (0,0) {$\mathbb{I}_3$};
\node[below] at (1,0) {$X$};
\node[below] at (2,0) {$X^{2}$};
\node[left] at (0,1) {$Z$};
\node[left] at (0,2) {$Z^{2}$};

\end{tikzpicture}
\vspace{10pt}
\caption{Commuting classes in dimension 3: points $(p_1,p_2)\in\mathbb{Z}_3^2$ and lines through the origin representing $\mathcal{C}_1,\dots,\mathcal{C}_4$. This geometric structure is a form of \textit{finite affine plane.}}
\label{fig:affine3D}
\end{figure}
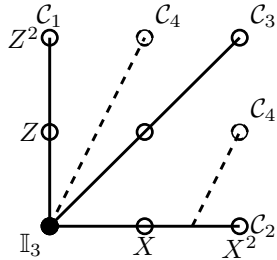

If we further abstract and represent the WH operators in terms of their exponents, which constrain the partitioning, 
\begin{equation}
\begin{aligned}
\mathcal{C}_{1} &= \{ (0,1),(0,2) \} \\
\mathcal{C}_{2} &= \{ (1,0), (2,0) \} \\
\mathcal{C}_{3} &= \{ (1,1), (2,2) \} \\
\mathcal{C}_{4} &= \{ (1,2), (2,1) \},
\end{aligned}
\end{equation}
a geometric representation in the form of a \textit{finite affine plane} \cite{Wootters2006} appears, as shown in Figure \ref{fig:affine3D}.

\begin{figure}[h]
\centering
\begin{tikzpicture}[scale=1.00, every node/.style={font=\small}]
  \foreach \x in {0,1,2,3} \foreach \y in {0,1,2,3}{
    \filldraw[white, draw=black, line width=0.6pt] (\x,\y) circle (2.3pt);
  }
  \fill (0,0) circle (2.6pt);
 \node at (2,0) {\LARGE $\times$};
 \node at (2,2) {\LARGE $\times$};

  \draw[line width=1.1pt] (0,0)--(0,3) node[above] {$\mathcal C_{1}$};
  \draw[line width=1.1pt] (0,0)--(3,0) node[right] {$\mathcal C_{2}$};
  \draw[line width=1.1pt] (0,0)--(3,3) node[above right] {$\mathcal C_{3}$};
  \draw[red, dashed, line width=1.1pt] (0,0)--(3/2,3) node[above] {$\mathcal C_{4}$};
  \draw[red, dashed, line width=1.1pt] (2,0)--(3,2) node[above right] {$\mathcal C_{4}$};
  \draw[blue, dotted, line width=1.1pt] (0,0)--(1,3) node[above left] {$\mathcal C_{5}$};
  \draw[blue, dotted, line width=1.1pt] (4/3,0)--(7/3,3) node[above left] {$\mathcal C_{5}$};
  
\node[below left=-1pt] at (0,0) {$\mathbb{I}_4$};
\node[below] at (1,0) {$X$};
\node[below] at (2,0) {$X^{2}$};
\node[below] at (3,0) {$X^{3}$};
\node[left] at (0,1) {$Z$};
\node[left] at (0,2) {$Z^{2}$};
\node[left] at (0,3) {$Z^{3}$};

\end{tikzpicture}
\vspace{10pt}
\caption{Commuting classes in dimension 4. The intersection points are marked with \LARGE $\times$.}
\label{fig:affine4D}
\end{figure}
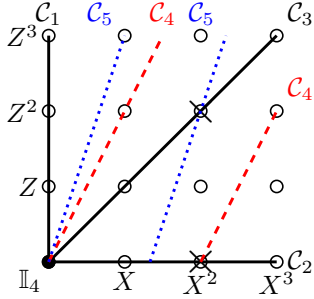

Notice that each class corresponds to a distinct, disjoint lines. In dimension $4$, on the other hand, employing WH operators in the similar manner yields classes that intersect each other, as shown in Figure \ref{fig:affine4D}. Therefore, maximal number $d+1=5$ of MUBs in dimension $4$, and in any prime-power dimension for that matter, cannot be constructed this way.
The underlying geometric structure relies on the fact that $\mathbb{Z}_3$ is a finite field, whereas $\mathbb{Z}_4$ is not. Consider two lines over an algebraic structure $\mathbb{F}$ equipped with addition and multiplication, $L_v = \{ av \ | \ a \in \mathbb{F}; \ v \in \mathbb{F}^2 \}$ and $L_w = \{ bw \ | \ b \in \mathbb{F}; \ w \in \mathbb{F}^2 \}$, with $v,w \in \mathbb{F}^2$. If the lines share a nonzero point, $av = bw$, then, provided every nonzero element of $\mathbb{F}$ admits a multiplicative inverse, one may cancel to obtain $v = (a^{-1} b) w$, implying that the lines coincide. Thus, distinct lines intersect only at the origin if and only if every nonzero element has an inverse, i.e., if $\mathbb{F}$ is a field.

In prime dimensions, the ring of integers modulo $d$, $\mathbb{Z}_d$, satisfies this property and therefore forms a finite field, whereas in composite dimensions such as $d=4$, $\mathbb{Z}_4$ contains noninvertible elements and the argument breaks down.

\subsubsection{\textbf{Prime-power dimensions}} In prime-power dimensions $d=p^m$, Bandyopadhyay et al. \cite{Bandyopadhyay2002} defines unitary operators acting on the composite Hilbert space $\mathcal{H}^{p^{m}} = \bigotimes_{i=1}^{m} \mathcal{H}^{p}$,
\begin{equation} \label{MUBformingPrimePower}
\begin{aligned}
U = \bigotimes_{j=1}^{m} M_{j}
\end{aligned}
\end{equation}
where
\begin{equation}
\begin{aligned}
M_j=X_p^{k_j} Z_p^{l_j}
\end{aligned}
\end{equation}
with $X_p$ and $Z_p$ denote the generalized Pauli operators in prime dimension $p$ and $k_j, l_j \in \mathbb{F}_p$ finite field of order $p$. The set of such operators form an orthogonal basis of unitary operators. Moreover, the commutation relation $\left[ U, U' \right] = 0$ yields the constraint $\sum_{j=1}^{m} k_{j} l'_{j} - \sum_{j=1}^{m} k'_j l_{j} =0 \ (\text{mod} \ p)$.

Following the representation introduced above, one may denote these operators in terms of their exponents and define $U \equiv (k_1,\cdots,k_m | l_1, \cdots, l_m) \equiv (\alpha|\beta) \in \mathbb{F}_p^{2m}$. And a set of $t$ such unitaries $\{ (\alpha_1|\beta_1),(\alpha_2|\beta_2),\cdots, (\alpha_t|\beta_t) \}$ is to be represented by a $t \times 2m$ matrix
\begin{equation}
\begin{aligned}
    \left(
    \begin{array}{c|c}
    \alpha_1 & \beta_1 \\
    \vdots & \vdots \\
    \alpha_t & \beta_t
    \end{array} 
    \right)
    \equiv
    \left(
    \begin{array}{c|c}
    k_{1,1}, ..., k_{1,m} & l_{1,1}, ..., l_{1,m} \\
    \vdots & \vdots \\
    k_{t,1}, ..., k_{t,m} & l_{t,1}, ..., l_{t,m} \\
    \end{array}
    \right),
\end{aligned}
\end{equation}
where each row corresponds to a unitary operator in the given set. It is further defined that the basis for each of the $p^m+1$ disjoint classes as $m \times 2m$ matrices
\begin{equation} \label{basisClass}
\begin{aligned}
\left(0_m | \mathbb{I}_m \right), \ \left(\mathbb{I}_m | 0_m \right), \  \left(\mathbb{I}_m | \mathbb{I}_m \right), \ \left(\mathbb{I}_m | A_3 \right), \ \cdots, \ \left(\mathbb{I}_m | A_{p^m} \right)
\end{aligned}
\end{equation}
corresponding to the classes $\mathcal{C}_0, \cdots, \mathcal{C}_{p^m}$. Therefore finding the maximal number of MUBs is reduced to finding the $A$-matrices. The pair of such matrices, to form relevant classes, have to obey the following two conditions:
\begin{itemize}
    \item \textit{\textbf{Commutativity:}} $A_{k'k}-A_{kk'}=0$.
    \item \textit{\textbf{Disjointness:}} $\text{det}(A_j-A_k) \neq 0$.
\end{itemize}

In dimension $d=4$, for example, the $A$-matrices take the form
\begin{equation} 
\begin{aligned}
A_1 = \begin{pmatrix} 
0 & 0 \\ 
0 & 0 \\
\end{pmatrix}, \ 
A_2 = \begin{pmatrix} 
1 & 0 \\ 
0 & 1 \\
\end{pmatrix}, \
A_3 = \begin{pmatrix} 
0 & 1 \\ 
1 & 1 \\
\end{pmatrix}, \ 
A_4 = \begin{pmatrix} 
1 & 1 \\ 
1 & 0 \\
\end{pmatrix}.
\end{aligned}
\end{equation}
This leads to the complete set of classes
\begin{equation} 
\begin{aligned}
&\mathcal{C}_{1} = \{ Z \otimes \mathbb{I}, \mathbb{I} \otimes Z, Z \otimes Z \} \\
&\mathcal{C}_{2} = \{ X \otimes \mathbb{I}, \mathbb{I} \otimes X, X \otimes X \} \\
&\mathcal{C}_{3} = \{ XZ \otimes \mathbb{I}, \mathbb{I} \otimes XZ, XZ \otimes XZ \} \\
&\mathcal{C}_{4} = \{ X \otimes Z, Z \otimes XZ, XZ \otimes XZ^2 \} \\
&\mathcal{C}_{5} = \{ XZ \otimes Z, Z \otimes X, XZ^2 \otimes XZ \}.
\end{aligned}
\end{equation}


\section{Generalized Numerical Construction} \label{section:construction}

\subsection{Gram Matrix}
The defining feature of MUBs lies in their inner-product relations. Since this feature is entirely encoded in pairwise overlaps between quantum states, it is natural to formulate the problem at the level of the \textit{Gram matrix}. For a set of quantum states $\{ \ket{\psi_{i}} \}_i$, the Gram matrix is defined by
\begin{equation} 
\begin{aligned}
 G_{i,j} = \braket{\psi_i}{\psi_j}.
\end{aligned}
\end{equation}

Consider, in dimension $d$, a set of $m$ MUBs $\Big\{ \mathcal{B}_\mu=\{\psi_{i}^{\mu} \}  \Big\}$ with $1 \leq i \leq d$ and $1 \leq \mu \leq m$. Each set $\mathcal{B}_\mu$ contains $d$ states, and there are $m$ such sets. We arrange all basis states into a $dm \times d$ matrix (with the proper normalization)
\begin{equation} 
\begin{aligned}
V= \frac{1}{\sqrt{m}} \begin{pmatrix} \bra{\psi_1^1} \\ \vdots \\ \bra{\psi_d^m}
\end{pmatrix}.
\end{aligned}
\end{equation}
The associated $dm \times dm$ Hermitian MUB Gram matrix is given by
\begin{equation} 
\begin{aligned}
G = VV^{\dagger}.
\end{aligned}
\end{equation}
In the index form, the MUB Gram matrix is
\begin{equation} \label{mubGramGeneral}
\begin{aligned}
(G^{\mu,\nu})_{i,j} = \frac{\braket{\psi_i^\mu}{\psi_j^\nu}}{m} = \frac{1}{m} \begin{cases}
\delta_{i,j} , & \mu=\nu; \\
e^{i \phi_{ij}^{\mu \nu}}/\sqrt{d}, & \mu \neq \nu \ \& \ j \nu > i \mu; \\
e^{-i \phi_{ij}^{\mu \nu}}/\sqrt{d}, & \mu \neq \nu \ \& \ j \nu < i \mu,
\end{cases},
\end{aligned}
\end{equation}
where $1 \leq i,j \leq d$ and $1 \leq \mu, \nu \leq m$ and $G^{\mu,\nu}$ is a block matrix with $G^{\mu,\mu} = \mathbb{I}_d$. The phases $\phi_{ij}^{\mu \nu}$ are the \textit{relative phases}, encoding the free parameters that determine the geometric structure of the corresponding MUBs. 

A crucial structural property of the MUB Gram matrix is that it is a projection operator: $G^2 = VV^{\dagger} VV^{\dagger} = G$, since $V^{\dagger}V = \mathbb{I}_d$. As such, it has the following features:
\begin{itemize}
    \item The eigenvalues are either $0$ or $1$.
    \item $G=G^2=G^3=\cdots=G^k$.
    \item $\text{Tr}(G) = \text{Tr}(G^k) = \text{rank}(G)= \text{dim}(\mathcal{H}) =  d$.
\end{itemize}
Moreover, the Gram matrix determines MUBs up to a unitary equivalence \cite{Waldron2018}. If two MUB configurations $\{\ket{\psi_i}\}$ and $\{\ket{\phi_i}\}$ are related by a global unitary transformation $U$, i.e., $U\ket{\psi_i} = \ket{\phi_i}$, then their Gram matrices are equivalent: $G_\phi=\braket{\phi_i}{\phi_j} = \langle \psi_i | U^{\dagger} U | \psi_j \rangle = \braket{\psi_i}{\psi_j} = G_\psi$. Thus, a MUB Gram matrix is invariant under global unitary transformation. Therefore, by encoding all the relative phases of a MUB configuration within a single Hermitian matrix, the Gram matrix provides a unitary-invariant and basis independent representation of the MUBs \cite{Waldron2018, samuel2024}.


\subsection{Trace Functions}

Any valid MUB Gram matrix G is a Hermitian projection operator. This imposes strong algebraic restrictions on the admissible relative phases that encode the geometry of the MUBs in Hilbert space. In particular, we can extract these restrictions through scalar constraints obtained from the $k$-th order traces $\text{Tr}(G^k)$. In what follows, we construct explicit trace-based scalar functions that allow us to restrict the numerical search space for MUBs.

The first and second order traces are trivial, providing no new information about the relative phases as the phases are canceled out: $\text{Tr}(G)=\sum_i |\langle \psi_i | \psi_i \rangle|$ and $\text{Tr}(G^2) = \sum_{ik} |\langle \psi_i | \psi_k \rangle|^2$. The third order trace, on the other hand, imposes the lowest order nontrivial constraint on the relative phases. We define $f_{d}(\Phi) \equiv \text{Tr}(G^3)$, expansion of which yields 
\begin{equation} \label{thirdOrderf}
\begin{aligned}
f_d(\Phi) \equiv \frac{(3m-2)d}{m^2}+\frac{6}{d^{3/2} m^3} \sum_{i,k,l=1}^{d} \sum_{\mu < \nu < \rho}^{m} \cos{(\phi_{\chi(\mu,i),\chi(\nu,k)} + \phi_{\chi(\nu,k),\chi(\rho,l)} - \phi_{\chi(\mu,i),\chi(\rho,l)} )},
\end{aligned}
\end{equation}
where $\Phi$ denotes the collection of relative phases and the index map $\chi(\mu,i) = (\mu-1)d+i$ flattens the double index.

It remains to determine how many trace constraints must be imposed. From the conditions $\text{Tr}(G^2)=\text{Tr}(G^3)=\text{Tr}(G^4)=d$ it follows that the relation $\sum_k \lambda_k^2 = \sum_k \lambda_k^3 = \sum_k \lambda_k^4=d$ implies $\sum_k \lambda_k^4 - 2 \sum_k \lambda_k^3 + \sum_k \lambda_k^2= \sum_k \lambda_k^2 (\lambda_k-1)^2=0$ which is true if and only if $\lambda_k \in \{0,1\}$. Therefore, under these trace conditions, $G$ is a projection operator of rank $d$. We conclude $\text{Tr}(G^3)=d$ and $\text{Tr}(G^4)=d$ are sufficient conditions for a valid projection matrix and a valid MUB Gram matrix in particular.


The last constraint function, therefore, is the function $g_{d}(\Phi) \equiv \text{Tr}(G^4)$ and it expands as
\begin{equation} \label{thirdOrderg}
\begin{aligned}
g_d(\Phi) &= \frac{1 - m + 5 d - 10 m d + 6 m^2 d}{m^3} + \frac{8}{m^4 d^2} \Bigg( \\
& \sum_{i,j,k,l=1}^{d} \sum_{\mu < \nu < \rho < \gamma}^{m}\Bigg[ \cos\left(\phi_{\chi(\mu,i),\chi(\nu,j)} + \phi_{\chi(\nu,j),\chi(\rho,k)} + \phi_{\chi(\rho,k),\chi(\gamma,l)} - \phi_{\chi(\mu,i),\chi(\gamma,l)}\right) \\
&\quad + \cos\left(\phi_{\chi(\mu,i),\chi(\nu,j)} + \phi_{\chi(\nu,j),\chi(\gamma,l)} - \phi_{\chi(\rho,k),\chi(\gamma,l)} - \phi_{\chi(\mu,i),\chi(\rho,k)}\right) \\
&\quad + \cos\left(\phi_{\chi(\mu,i),\chi(\rho,k)} - \phi_{\chi(\nu,j),\chi(\rho,k)} + \phi_{\chi(\nu,j),\chi(\gamma,l)} - \phi_{\chi(\mu,i),\chi(\gamma,l)}\right) \Bigg] \\
& + \sum_{j,l=1}^{d}\sum_{1 \le i < k \le d} \sum_{\nu < \mu < \gamma}^{m} \cos\left(-\phi_{\chi(\nu,j),\chi(\mu,i)} + \phi_{\chi(\nu,j),\chi(\mu,k)} + \phi_{\chi(\mu,k),\chi(\gamma,l)} - \phi_{\chi(\mu,i),\chi(\gamma,l)}\right) \\
& + \sum_{j,l=1}^{d}\sum_{1 \le i < k \le d} \sum_{\mu < \nu < \gamma}^{m} \cos\left(\phi_{\chi(\mu,i),\chi(\nu,j)} - \phi_{\chi(\mu,k),\chi(\nu,j)} + \phi_{\chi(\mu,k),\chi(\gamma,l)} - \phi_{\chi(\mu,i),\chi(\gamma,l)}\right) \\
& + \sum_{i,k=1}^{d}\sum_{1 \le j < l \le d} \sum_{\mu < \rho < \nu}^{m} \cos\left(\phi_{\chi(\mu,i),\chi(\nu,j)} - \phi_{\chi(\rho,k),\chi(\nu,j)} + \phi_{\chi(\rho,k),\chi(\nu,l)} - \phi_{\chi(\mu,i),\chi(\nu,l)}\right) \\
& + \sum_{1 \le i < k \le d}^{d} \sum_{1 \le j < l \le d} \sum_{\mu < \nu}^{m} \cos\left(\phi_{\chi(\mu,i),\chi(\nu,j)} - \phi_{\chi(\mu,k),\chi(\nu,j)} + \phi_{\chi(\mu,k),\chi(\nu,l)} - \phi_{\chi(\mu,i),\chi(\nu,l)}\right)
\Bigg).
\end{aligned}
\end{equation}

In this subsection, we developed the MUB Gram matrix and trace functions in the general form. In our numerical search, unless specified otherwise, we are concerned with the maximal number of MUBs, i.e., $m=d+1$ case.
\subsection{Construction}

To construct MUBs in the $d$-dimensional Hilbert space $\mathcal{H}^{d}$, we look for phase configurations $\Phi=\{\phi_{i,j}\}$ that simultaneously obey the trace constraints $f_{d}(\Phi)=d$ and $g_{d}(\Phi)$. For the numerical search, we define the \textit{constraint function} $F(\Phi) = \sqrt{\bigl(f_d(\Phi)-d\bigr)^2+ \bigl(g_d(\Phi)-d\bigr)^2}$ for which $F(\Phi)=0$ produces a perfect rank-$d$ MUB Gram matrix.

By first assigning random initial points to $d^2 m (m-1)/2$ relative phases in the set $\Phi$, the search is performed by minimizing the constraint function $F(\Phi)$ using the Wolfram Mathematica's built-in $\text{FindMinimum}$ function, iteratively updating the phase configuration. We used two-step minimization strategy: first by using conjugate gradient method then by using the quasi-Newton method. A configuration is accepted as a solution if the configuration is within the proper numerical tolerance we specify.


\subsection{Isolated and Continuous Families of Solutions} \label{section:vicinity}

The construction of generalized numerical MUBs reduces the problem to finding the proper relative phases $\Phi=\{\phi_{ij}^{\mu \nu}\}$ that satisfy the constraint function $F(\Phi)$. A valid MUB solution, represented by the set $\Phi$, is a point in the $d^2 m (m-1)/2$-dimensional phase space. The constraint function $F(\Phi)=0$ defines a manifold over this configuration space, and the MUB solutions correspond to the points on this manifold. Then a natural question to ponder is whether the solutions are isolated points or form a continuous family of solutions.

In order to answer this question, we need to investigate the neighbourhood of the solutions. A straightforward way to do that is to perturb the solutions while still obeying the constraint function and preserving the projection property of the MUB Gram matrix. A computationally simpler method operates at first order and was introduced in \cite{Bruzda2017} for MUBs as well as SIC-POVMs. This method first defines a unitary matrix $U = \mathbb{I} - 2G$, where $G$ is the MUB Gram matrix and considers perturbations of the form
 \begin{equation}
     V_{jk}(t) = U_{jk} e^{i t R_{jk}},
 \end{equation}
where $R$ is a real antisymmetric matrix parametrizing the perturbation and $t \in \mathbb{R}$. The problem reduces to determining how many independent parameters $R_{jk}$ can be introduced while preserving unitarity of $V(t)$ in the solution neighbourhood of $t=0$. Imposing the linearized unitarity condition $\frac{d}{dt}\left[V(t)\,V(t)^\dagger\right]_{t=0} = 0$ yields a real linear system whose rank $r$ counts the number of independent constraints.

The \textit{restricted defect} is then defined as $\Delta = \tau_{res} - r$, where $\tau_{res} = (N-1)(N-2)/2 - z$ is the total number of free perturbation parameters, with $z$ denoting the number of zero entries in the upper triangular part of $U$. If $\Delta = 0$, there are no non-trivial directions and the solution is isolated. If $\Delta > 0$, there exist non-trivial directions in the linearized problem, indicating that the solution may, but not necessarily, part of a continuous family; confirming a genuine family requires verifying that the linear perturbations do extend to exact solutions. 

To rigorously verify that a positive restricted defect corresponds to existence of a continuous family solutions, we have to explicitly solve for the phase values $R_{jk}$ while still obeying the MUB Gram matrix constraint. We may simply take the constraint as $\Psi(G) = G^2 - G = 0$ and look for perturbations $\delta G$ with which resulting Gram matrix still obeys this constraint in the first order approximation $\Psi(G+\delta G) \approx G \cdot \delta G + \delta G \cdot G - \delta G = 0$. Here we may consider the perturbed Gram matrix as simply $G_{jk} = G_{jk} e^{i t R_{jk}} \approx G_{jk} (1 + i t R_{jk})$, or equivalently as $G \approx G (\mathbb{I}_{dm} + i t R)$ for the antisymmetric matrix $R$ which has non-zero elements only in the positions corresponding to the positions of the relative phases in the original Gram matrix. In order to solve for the perturbation phases inscribed within $R$, we may reduce the problem to system of linear equations by defining a vector composed of the elements of the constraint relation $(\Psi(G+\delta G))_{jk}$.

Ultimately, we would end up with the problem of solving for the null space of the Jacobian of the constraint relation. The null space includes vectors corresponding to gauge phase shifts which are simple rephasings of the Gram matrix and do not change the MUB configuration at all. Excluding these gauge shifts, we would be left with vectors which span the tangent space at a solution. The number of such perturbations is actually what restricted defect counts for. By repeatedly marching forward along such tangent vectors one can trace a finite, continuous trajectory through the configuration space confined to the solution manifold.

\section{Group Theoretical Classification: Symmetries and Triple Products} \label{section:classification}

\subsection{Bargmann Invariants and Triple Product Tensor}

The physical content of a set of state vectors is encoded in their pairwise inner products. However, each quantum state is defined only up to an overall (global) phase: $\ket{\psi}$ and $e^{i\phi_k}\ket{\psi}$ represent the same physical state. To describe structural features of MUB configurations in a way that is independent of these phase choices, we use Bargmann invariants (also known as \textit{m-products}).

Given a set of states $\{ \ket{\psi_{i}} \}_i^{N}$, the $m$-th order Bargmann invariant associated with indices $1 \leq j_1,j_2,\cdots,j_m \leq N$, is defined as 
\begin{equation}
\Delta_m(j_1, j_2, \dots, j_m) = \braket{\psi_{j_1}} {\psi_{j_2}} \braket{\psi_{j_2}}{\psi_{j_3}} \cdots \braket{\psi_{j_m}}{\psi_{j_1}}.
\end{equation}
Besides unitary transformations, these quantities remain invariant under \textit{projective unitary transformations}, $\ket{\psi_k} \to c_k U \ket{\psi_k}$ with $|c_k|=1$ and $U$ is unitary \cite{ChienYowWaldron2016, ChienWaldron2018}.

For our purposes, the lowest-order nontrivial phase-sensitive invariant is the triple product which encodes the relative phase information that distinguishes different MUB structures ($m=3$):
\begin{equation}
\begin{aligned}
T_{ijk} \equiv \Delta_3(i,j,k) = \braket{\psi_{i}}{\psi_{j}} \braket{\psi_{j}}{\psi_{k}} \braket{\psi_{k}}{\psi_{i}}.
\end{aligned}
\end{equation}
For a complete set of $d+1$ MUBs in dimension $d$, the indices range over $N=d(d+1)$ states, so $T_{ijk}$ forms an $N\times N\times N$ tensor. Moreover, if the Gram matrix is known, one can build the triple product tensor as $T_{ijk} = G_{ij}G_{jk}G_{ki}$ \cite{ApplebyFlammiaFuchs2009}. In particular, following the block structure of the MUB gram matrix in Equation \eqref{mubGramGeneral}, flattening the Gram matrix in the form $G_{I,J}$ with the index transformation $I(\mu,i) = (i-1)d+\mu$ and $J(\nu,j) = (j-1)d+\nu$, the definition of the triple product tensor can be utilized as $T_{IJK}$.


Two MUB sets that are projectively equivalent have identical triple product tensors, whereas inequivalent constructions produce distinct set of phases in the triple product tensor \cite{samuel2024, ChienYowWaldron2016}. Thus the triple product tensor serves as a suitable tool for MUB classification. Following Samuel and Gedik \cite{samuel2024}, we denote the set of distinct phases appearing in the triple product tensor as \textit{generating set}: $\text{gen}(T_{ijk})=\{\phi_{ijk}\}$, where $\phi_{ijk}=\text{arg}(T_{ijk})$.

If two MUBs have different generating sets, they are necessarily inequivalent. However, the converse does not automatically hold: two constructions sharing the same generating set are not guaranteed to be equivalent. In such cases, one must verify whether their Gram matrices are related by a permutation (i.e., there must exist a permutation matrix $U$ such that $G' = U^\dagger G U$). If so, the two configurations are said to be \textit{isomorphic} \cite{samuel2024}.



\subsection{Permutation Symmetry}

To characterize the group theoretic properties of MUBs, we investigate their symmetries. While the algebraic structures of standard analytical constructions are well-understood, generalized numerical MUBs may, in principle, yield new forms of solutions since no \textit{a priori} assumption of an algebraic structure is made in their construction.

We employ the triple product tensor to express the gauge-invariant phase relations of the MUB set \cite{samuel2024,ChienYowWaldron2016,ApplebyFlammiaFuchs2009}. The underlying group structure is identified through the automorphism group of this tensor, defined as the set of permutations $\sigma$ that preserve the tensor elements:
\begin{equation} \label{automorphismGroup}
\begin{aligned}
\text{Aut}(T_{ijk}) = \left\{ \sigma \ | \ T_{\sigma(i)\sigma(j)\sigma(k)} = T_{ijk} \right\},
\end{aligned}
\end{equation}
where $1 \leq i,j,k \leq dm$. As the triple products encode the geometry of the MUBs, it is a natural representation with which we can compute the symmetries of a MUB structure. In our calculations, for computational efficiency, we slice the triple product tensor into $d(d+1) \times d(d+1)$ matrices by keeping the first index fixed, $T_{jk}^{(a)}=\braket{\psi_{a}}{\psi_{j}} \braket{\psi_{j}}{\psi_{k}} \braket{\psi_{k}}{\psi_{a}}$, investigate the symmetries of each slice, and then identify the symmetries common to them all: $\text{Aut}(T_{ijk}) = \left\{ \sigma \ | \ T_{\sigma(j)\sigma(k)}^{(a)} = T_{jk}^{(a)} \ \forall a \right\}$. These permutations are simply relabelings and repositionings of the quantum states among and within the bases, preserving the geometry of the MUBs.

\subsection{Symmetry Group of the Known Solutions} \label{section:symmetryKnownSols}

In the dimensions $3,4$ and $5$ we carry out our investigations for the full set of MUBs in this paper, the known analytical solutions of complete set of MUBs are all equivalent \cite{Brierley2009, Blanchfield2014}. Therefore, we shall restrict our scope to the symmetries related to these analytical constructions (we have already discussed one form of such constructions in Section \ref{section:knownSols}). 

Considering the standard construction using WH operators, MUBs are generated by the eigenstates of the WH operators. The symmetry group of the MUBs is then related to the symmetries of the underlying algebraic structure of the WH operators. In particular, any unitary that permutes the WH operators under conjugation 
will permute the maximal abelian subgroups of $\textbf{H}(d)$ among themselves, and 
hence permute their joint eigenbases — that is, the MUB bases — among themselves. 
The group of all such unitaries is precisely the \textit{normalizer} of the WH group, which is known as the Clifford group.

\subsubsection{\textbf{Clifford group: the normalizer:}} The Clifford group is defined as the normalizer of the WH group, i.e., the set of unitaries that preserve the WH group under conjugation \cite{ScottGrassl2010,Appleby2005}:
\begin{equation}
\begin{aligned}
\mathcal{\textbf{C}}(d)=\{U \ | \ U \textbf{H}(d) U^\dagger = \textbf{H}(d)\}.
\end{aligned}
\end{equation}
In other words, each Clifford group element permutes and rephases the WH operators:
\begin{equation}
  U D_{\textbf{p}} U^{\dagger} = e^{i g(\textbf{u})} D_{f(\textbf{u})}.
\end{equation}
The precise form of the functions $f$ and $g$, therefore the nature of the Clifford group, depends on the algebraic structure of the phase space we are working on — the algebraic structure $\textbf{u}$ belongs to. This distinction proves more so crucial in investigating the symmetries of standard construction we introduced in Section \ref{section:knownSols}. In prime dimensions, the phase space was simply formed by a ring of integers $\mathbb{Z}_d$, as in $\textbf{p} \in \mathbb{Z}_d^2$; while in prime-power dimensions, the phase space was formed by a finite field $\mathbb{F}_d$. Below we follow the notation that Appleby introduced in \cite{Appleby2005} and briefly overview \textit{unrestricted Clifford group} and \textit{restricted Clifford group}.

\subsubsection*{\textbf{Unrestricted Clifford group:}} 

The elements of the Clifford group induce a \textit{symplectic transformation} on the WH displacement operators, i.e., for each element $U$, there exist a matrix $F$ and a vector $\chi$ such that
\begin{equation}
  U D_{\textbf{p}} U^{\dagger} =\omega^{ \langle \chi, F \textbf{u} \rangle } D_{F \textbf{u}},
\end{equation}
where $\omega = e^{i 2 \pi/d}$ and $\langle \textbf{p}, \textbf{q} \rangle = q_1 p_2 - q_2 p_1$. The unrestricted Clifford group is defined such that the algebraic structure of the phase space is a ring of integers $\mathbb{Z}_d$, regardless of whether $d$ is prime or not. Therefore, with $\textbf{u} \in \mathbb{Z}_d^2$ and $\chi \in \mathbb{Z}_d^2$. The matrix $F$ belongs to the \textit{special linear group} defined as
\begin{equation}
\begin{aligned}
SL(2,\mathbb{Z}_{\overline{d}})=\left\{ F= \begin{pmatrix}
a & b \\
c & d
\end{pmatrix} \ | \ \text{det}(F)=1 \  (\text{mod } \overline{d}) \right\},
\end{aligned}
\end{equation}
where $\overline{d} = d$ if $d$ is odd and $\overline{d} = 2d$ if $d$ is even \cite{Appleby2005}. 

It is proven \cite{ScottGrassl2010,Appleby2005} that for each pair $(F,\textbf{u})$ there is a corresponding operator $U \in \textbf{C}(d)$. In fact, each such pair constitutes the group $\textbf{c}(d) = \{ (F | \textbf{u}) \ | \ F \in SL(2,\mathbb{Z}_{\overline{d}} ), \ \textbf{u} \in \mathbb{Z}_d^2 \}$ which has the isomorphism relation in \emph{odd} dimensions as
\begin{equation} \label{CliffordGroupIsomorphism}
  \textbf{c}(d) \cong SL(2,\mathbb{Z}_{d}) \rtimes \mathbb{Z}_d^{2}.
\end{equation}
In even dimensions, on the other hand, $\textbf{c}(d)$ is isomorphic to a quotient group of $SL(2,\mathbb{Z}_{2d}) \rtimes \mathbb{Z}_d^{2}$ \cite{ScottGrassl2010,Appleby2005}. The unrestricted Clifford group in this form captures only the action of its elements, discarding all the global phases that may appear in the unitaries. Namely, $\textbf{c}(d) = \textbf{C}(d)/\textbf{I}(d)$, where $\textbf{I}(d)$ is a group consisting of all the identity operators $e^{i\zeta}\mathbb{I}_d$. As the global phases are irrelevant for us, Clifford group in this form will be our main object of interest (we shall refer to this group as the Clifford group). In contrast to $\textbf{C}(d)$ in which unitaries with the same action on the displacement operators appear repeatedly with different global phases, in $\textbf{c}(d)$ they are ruled out. 

The order of the unrestricted Clifford group in this form is, for any dimension $d$, regardless of being odd or even \cite{Appleby2005}, given by
\begin{equation}
\begin{aligned}
|\mathcal{\textbf{c}}(d)| = |SL(2,\mathbb{Z}_d) \rtimes \mathbb{Z}_d^{2}|=d^2|SL(2,\mathbb{Z}_d)|.
\end{aligned}
\end{equation}
In prime dimensions $d=p$, the order of the special linear group is $|SL(2,\mathbb{Z}_p)|=p(p^2-1)$, hence the order of the unrestricted Clifford group reduces to \cite{Appleby2005}
\begin{equation} \label{sizeCliffordPrimeDim}
\begin{aligned}
|\textbf{c}(p)|=p^3(p^2-1).
\end{aligned}
\end{equation}

\subsubsection*{\textbf{Restricted Clifford group:}} 

The restricted Clifford group, $\textbf{C}^{\text{res}}(d)$, translates the definition of the unrestricted one with the utilization of the finite field $\mathbb{F}_d$ (i.e., $\textbf{u} \in \mathbb{F}_d^2$) rather than ring of integers, and it is the subgroup of the unrestricted Clifford group \cite{Appleby2009}. Since in prime dimensions ring is a field, the restricted Clifford group is equivalent to the unrestricted one. The matrix $F$ that represents the symplectic transformation is now rather belongs to $SL(2,\mathbb{F}_d)$. The isomorphism relation in a odd dimensions is now given by \cite{Zhu2015}
\begin{equation} \label{RestrictedCliffordGroupIsomorphism}
  \textbf{c}^{\text{res}}(d) \cong SL(2,\mathbb{F}_{d}) \rtimes \mathbb{F}_d^{2}.
\end{equation}
We expect that the order of the restricted Clifford group is also given by $|\textbf{c}^{\text{res}}(d)| = d^2 |SL(2,\mathbb{F}_d)|$ for any dimension $d$, which reduces to $|\textbf{c}^{\text{res}}(d)|=d^3(d^2-1)$ in prime-power dimensions.

\subsubsection*{\textbf{Extended Clifford group:}} The Clifford group we introduced thus far covers only the unitaries that normalize WH group. Anti-unitaries also preserve the WH group under conjugation, i.e., for an anti-unitary $J$, we have
\begin{equation}
\begin{aligned}
J \textbf{H}(d) J^{\dagger} = \textbf{H}(d).
\end{aligned}
\end{equation}
With this additional symmetry, we define \textit{extended unrestricted Clifford group} which covers all the unitaries and anti-unitaries that normalize WH group:
\begin{equation}
\begin{aligned}
\textbf{ec}(d) = \{ (F | \textbf{p}) \ | \ F \in ESL(2,\mathbb{Z}_d), \ \textbf{p} \in \mathbb{Z}_d^2 \} \cong ESL(2,\mathbb{Z}_d) \rtimes \mathbb{Z}_d^2.
\end{aligned}
\end{equation}
The extended special linear group $ESL$ is defined as a group of $2 \times 2$ matrices
\begin{equation}
\begin{aligned}
ESL(2,\mathbb{Z}_{\overline{d}}) =\big\{ F \ | \ \text{det}(F)=\pm1 \  (\text{mod } \overline{d}) \big\}.
\end{aligned}
\end{equation}
The same argument applies to the restricted Clifford group. The order of the Clifford group, whether unrestricted or restricted, doubles in the extended form \cite{Appleby2005}:
\begin{equation} \label{sizeExtendedClifford}
\begin{aligned}
|\textbf{ec}(d)| = 2 |\textbf{c}(d)|.
\end{aligned}
\end{equation}

Within the construction of standard MUBs — where bases arise as joint eigenbases of maximal abelian subgroups of $\textbf{H}(d)$ — the Clifford group constitutes the symmetry group of the MUB structure \cite{Zhu2015}. Specifically, Chien and Waldron \cite{ChienWaldron2018} define the \textit{projective symmetry group} as the group of all index permutations $\sigma$ such that $\Delta_m(j_1,j_2, \cdots, j_m) = \Delta_m(j_{\sigma(1)},j_{\sigma(2)}, \cdots, j_{\sigma(m)})$. Applying this framework to the standard construction of $d+1$ MUBs in $\mathcal{H}^d$, their computational results for $d \leq 5$ show that every projective symmetry of the standard MUBs is accounted for by a Clifford element. This leads them to conjecture that for all primes $d$, the standard WH construction of $d+1$ MUBs has no projective symmetries other than those given by the Clifford group \cite{ChienWaldron2018}. 

To the verification of the permutation symmetry approach Equation \eqref{automorphismGroup}, for the standard MUBs in dimension $3$ and $5$ we find that the Clifford group, via its generators \cite{Morvan2021}, induces permutations on the $d(d{+}1)$ MUB states, and that the group generated by these permutations coincides exactly with the permutations obtained from the symmetries of the triple product tensor. This establishes that $\text{Aut}(T_{ijk})$ is equivalent to the action of the Clifford generators on the MUB states. In dimension $3$, the size of the automorphism group is $|\text{Aut}(T_{ijk})|=216$, which matches the size of the Clifford group in dimension $3$ (see Equation \eqref{sizeCliffordPrimeDim}). In dimension $5$, we find that $|\text{Aut}(T_{ijk})|=3000$, which also matches the size of the Clifford group in dimension $5$.

Moreover, the structural analysis using the \textit{Computational Discrete Algebra} (GAP) software identifies the automorphism group of the $3$-dimensional MUBs with the isomorphism relation $\text{Aut}(T_{ijk}) \cong SL(2,\mathbb{Z}_{3}) \rtimes \mathbb{Z}_3^{2}$, thus confirming our expectation. As for the $5$-dimensional case, the isomorphism $\text{Aut}(T_{ijk}) \cong SL(2,\mathbb{Z}_{5}) \rtimes \mathbb{Z}_5^{2}$ is also confirmed. The prime-power dimension $4$ case is more subtle. The structural analysis identifies the automorphism group of the $4$-dimensional case with the decomposition $\text{Aut}(T_{ijk}) \cong \mathbb{Z}_2^4 \rtimes S_5$. It is known that $S_5$ is an extension of $A_5$ by $\mathbb{Z}_2$: $S_5 \cong A_5 \rtimes \mathbb{Z}_2$, where $A_5$ is the alternating group of order $5$. Since the alternating group is isomorphic to the special linear group $A_5 \cong SL(2,\mathbb{F}_4)$, and $\mathbb{Z}_2^4 \cong \mathbb{F}_4^2$, we can further decompose the automorphism group as $\text{Aut}(T_{ijk}) \cong \mathbb{F}_4^2 \rtimes SL(2,\mathbb{F}_4) \rtimes \mathbb{Z}_2$. The order of the group, as such, becomes $|\text{Aut}(T_{ijk})| = 2 \ |\mathbb{F}_4^2 \rtimes SL(2,\mathbb{F}_4)| = |\text{\textbf{ec}}^{\text{res}}(4)|$. Hence, it coincides with the structure of the extended restricted Clifford group in dimension $4$.

\section{Results} \label{section:results}

\subsubsection*{\textbf{Dimensions 3, 4, 5:}} For the dimensions $3$, $4$, and $5$ in which the search was conducted, a validity threshold of $F(\Phi) < 10^{-20}$ is required. Additionally, we polished some solutions up to $100$ digits to confirm whether they are proper solutions. We observe that all the solutions behave well under this threshold, with $F(\Phi)$ values around $10^{-29}$ for dimension $3$, and around $10^{-28}$ for dimensions $4$ and $5$. The average computation time per solution grows by roughly a factor of $20$ from dimension $3$ to dimension $4$, and by another factor of roughly $14$ from dimension $4$ to dimension $5$. The success rate, per trial, drops significantly as the dimensions increases. Since the phase space expands with the dimension, the search becomes more difficult and time consuming.

Table \ref{tab:construction} summarizes the search results.

\begin{table}[h]
\centering
\begin{tabular}{@{}cccccc@{}}
\hline
$d$ & \textbf{No.\ of Solutions} & \textbf{Precision (digits)} & $F(\Phi)$ & \textbf{Avg.\ Time (s)} & \textbf{Success Rate} \\
\hline
$3$ & $1,000,000$ & $30$ & $\approx 10^{-29}$ & $\approx 0.06$ & $97\%$ \\
$4$ & $800,000$ & $30$ & $\approx 10^{-28}$ & $\approx 1.2$  & $75\%$ \\
$5$ & $100,000$ & $30$ & $\approx 10^{-28}$ & $\approx 17$ & $40\%$ \\
\hline
\end{tabular}
\caption{Summary of MUB Gram matrix search results in dimensions $3$, $4$, and $5$. Success rate quantifies the percentage of the found solutions out of the total number of search attempts.}
\label{tab:construction}
\end{table}

\begin{figure}[h]
\centering
\begin{subfigure}[t]{0.48\textwidth}
  \centering
  \includegraphics[height=4.0cm,keepaspectratio]{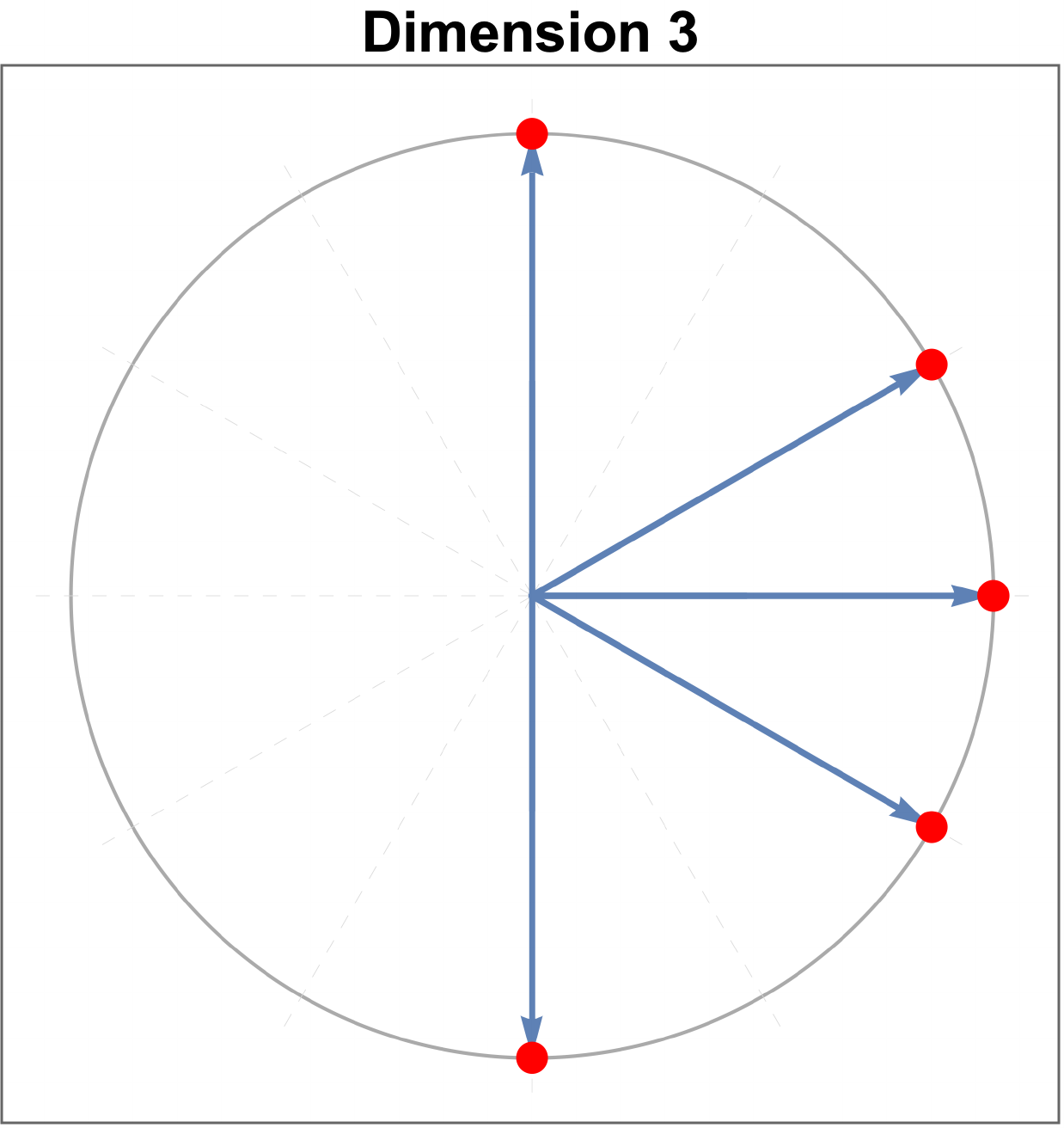}
  \caption{Unit circle, $d=3$: five distinct phases.}
  \label{fig:genSet3D_circle}
\end{subfigure}
\hfill
\begin{subfigure}[t]{0.48\textwidth}
  \centering
  \includegraphics[height=4.0cm,keepaspectratio]{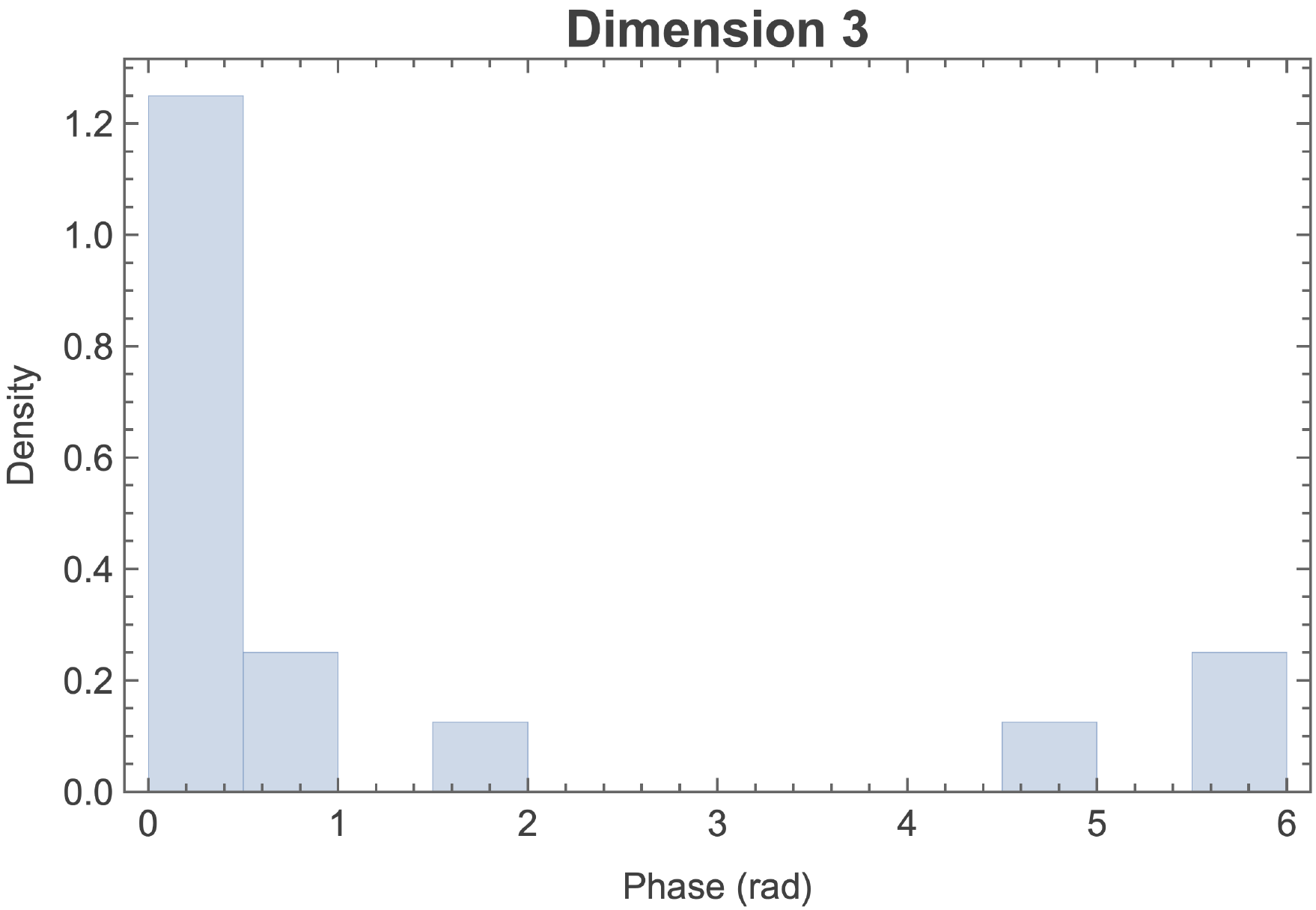}
  \caption{Phase distribution, $d=3$.}
  \label{fig:genSet3D_dist}
\end{subfigure}

\vspace{0.3cm}

\begin{subfigure}[t]{0.48\textwidth}
  \centering
  \includegraphics[height=4.0cm,keepaspectratio]{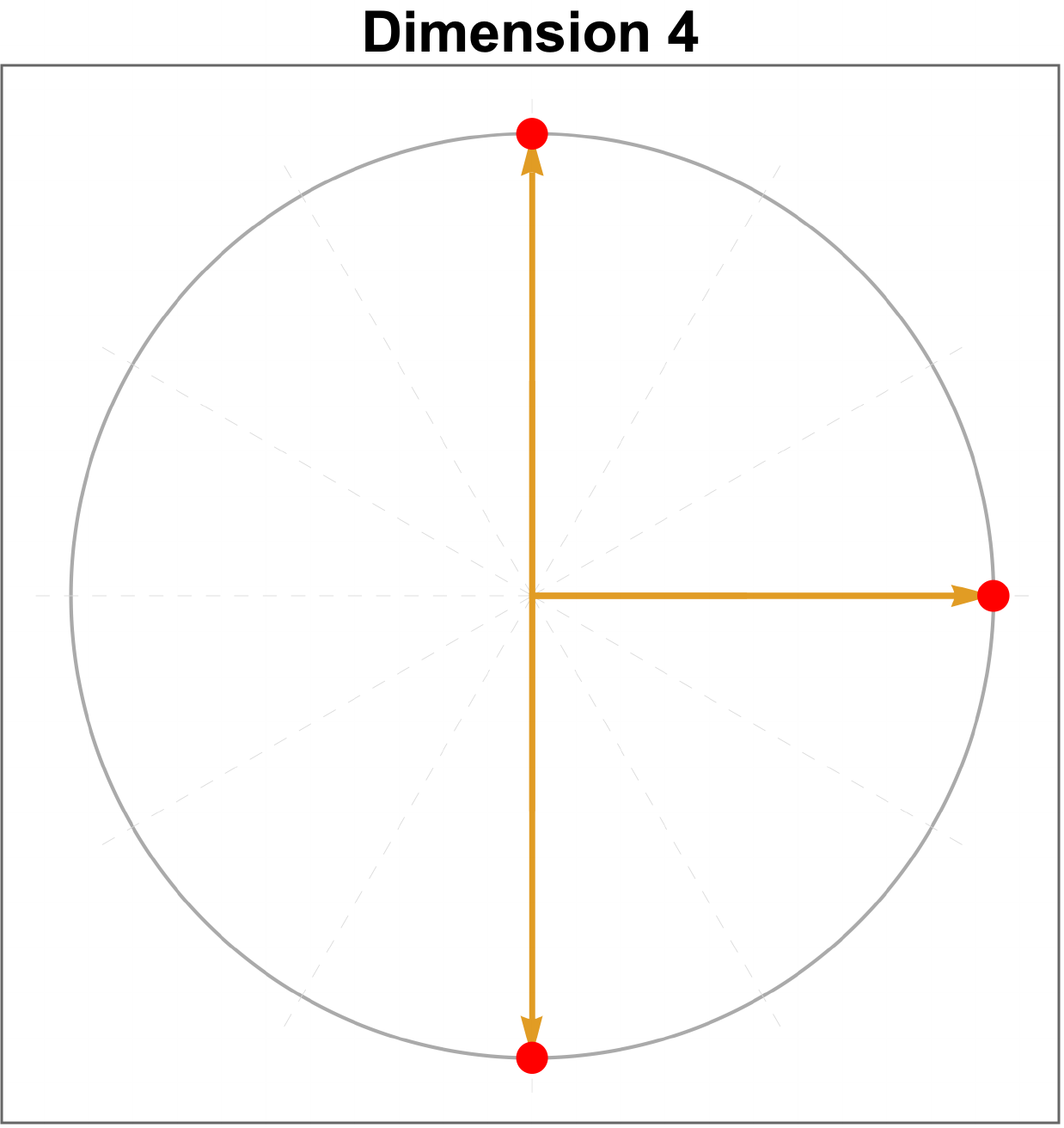}
  \caption{Unit circle, $d=4$: three distinct phases.}
  \label{fig:genSet4D_circle}
\end{subfigure}
\hfill
\begin{subfigure}[t]{0.48\textwidth}
  \centering
  \includegraphics[height=4.0cm,keepaspectratio]{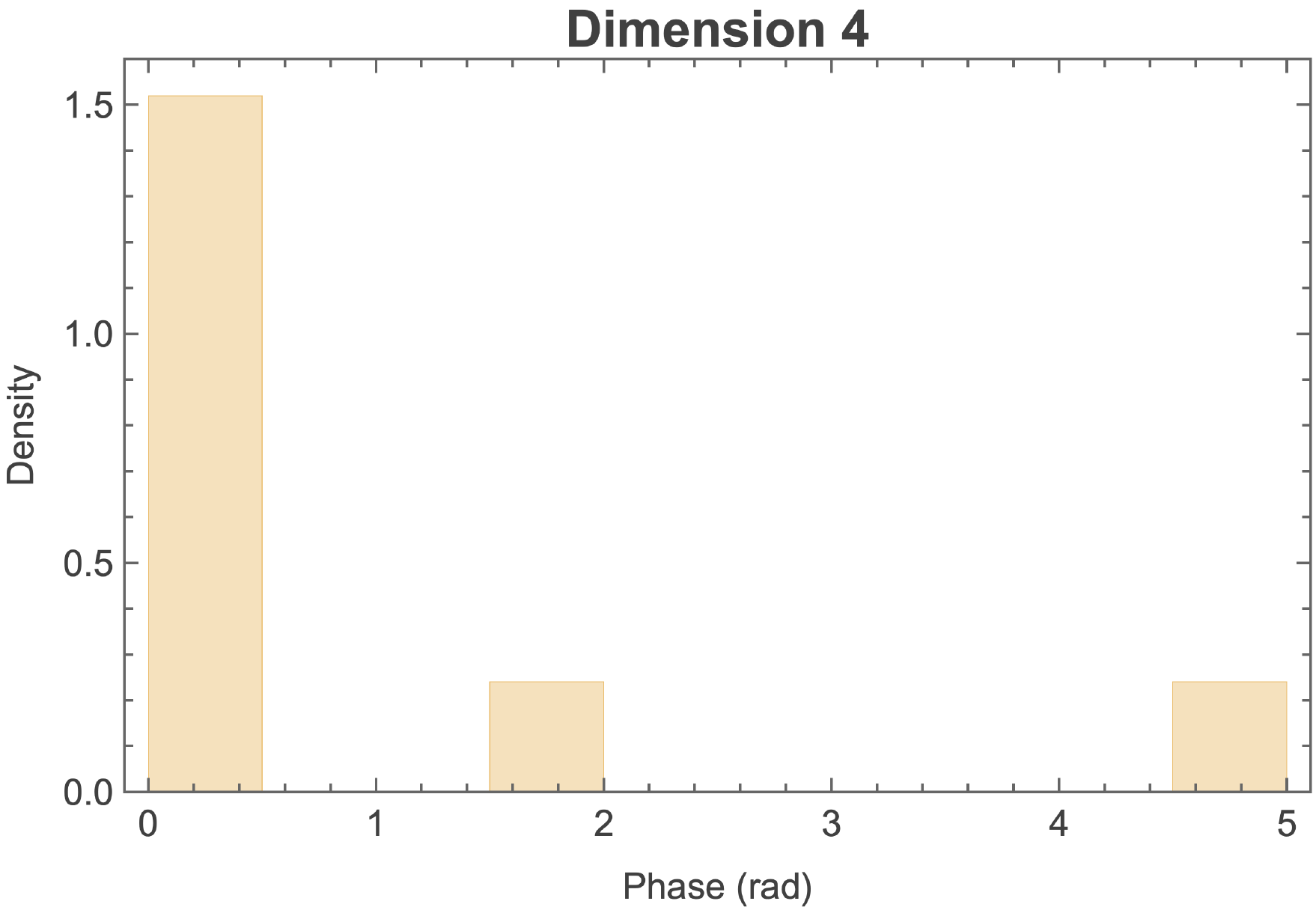}
  \caption{Phase distribution, $d=4$.}
  \label{fig:genSet4D_dist}
\end{subfigure}

\vspace{0.3cm}

\begin{subfigure}[t]{0.48\textwidth}
  \centering
  \includegraphics[height=4.0cm,keepaspectratio]{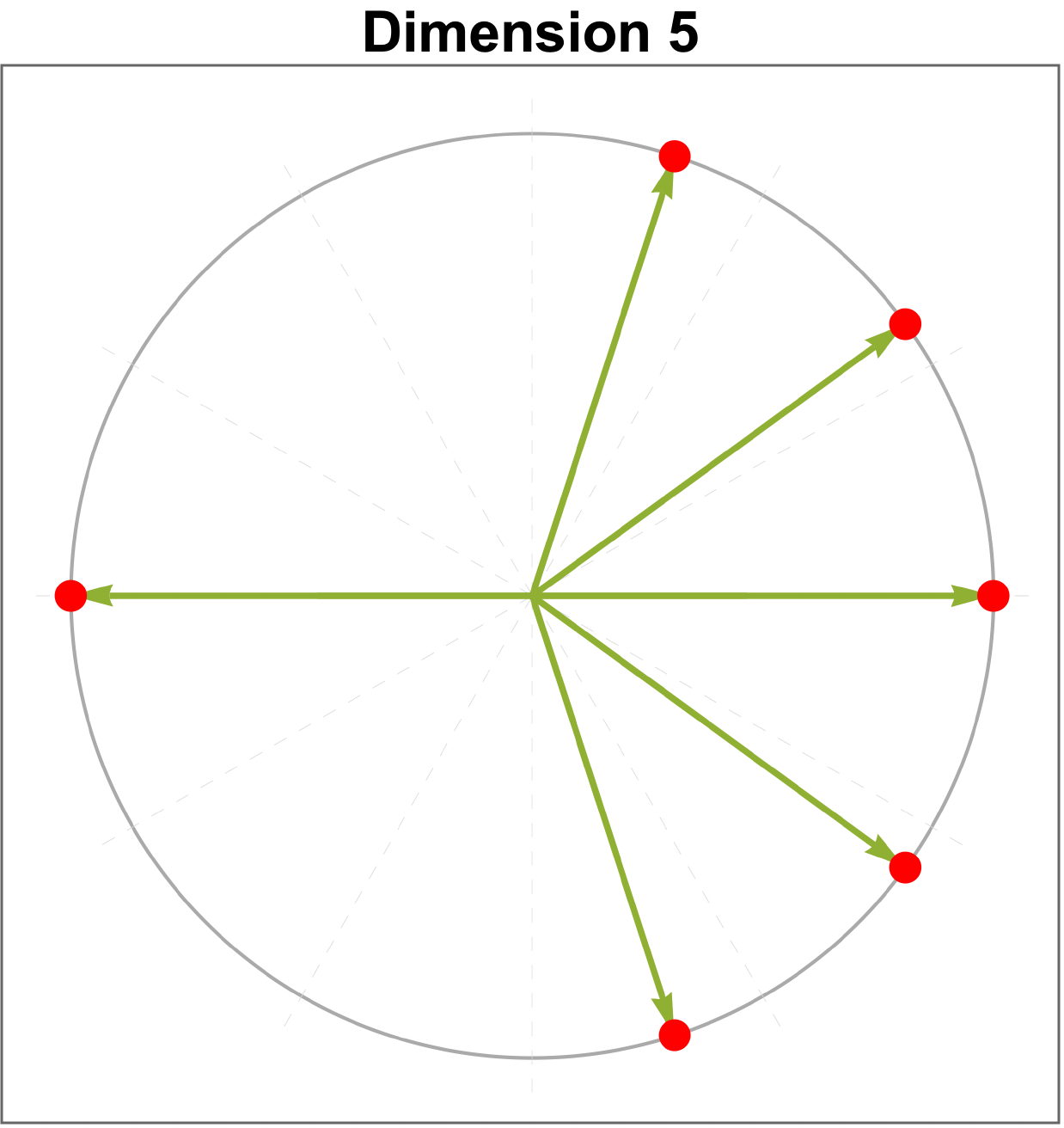}
  \caption{Unit circle, $d=5$: six distinct phases.}
  \label{fig:genSet5D_circle}
\end{subfigure}
\hfill
\begin{subfigure}[t]{0.48\textwidth}
  \centering
  \includegraphics[height=4.0cm,keepaspectratio]{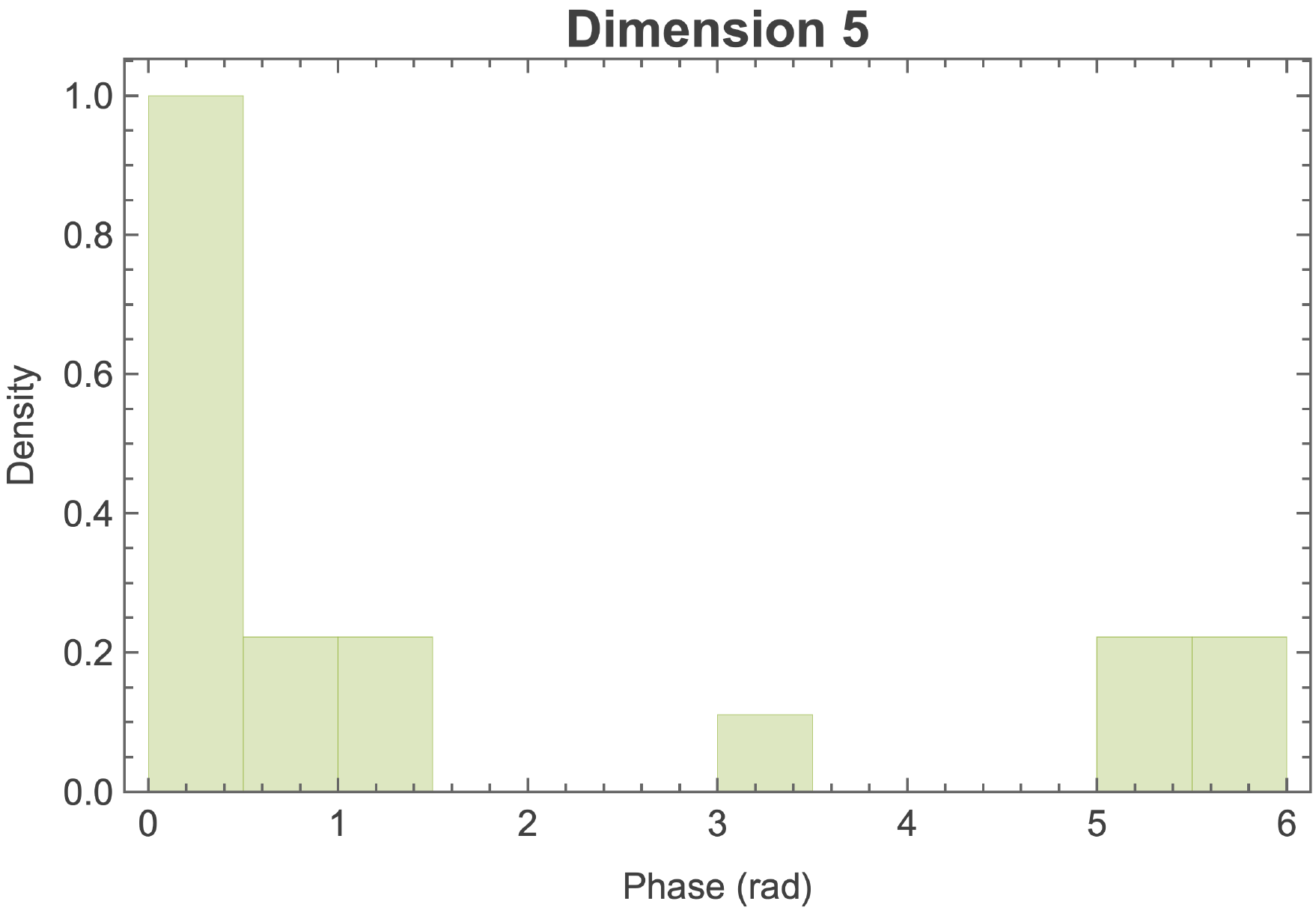}
  \caption{Phase distribution, $d=5$.}
  \label{fig:genSet5D_dist}
\end{subfigure}

\caption{Generating set elements of the triple product tensor plotted on the unit circle (left column) and their distributions in the histogram form (right column), for dimensions $d=3$ (top), $d=4$ (middle), and $d=5$ (bottom). Red dots mark the distinct phases.}
\label{fig:genSets_all}
\end{figure}

We list the characteristic properties of the solutions we found in Table \ref{tab:characterization}. Across all the solutions found in dimensions $3$, $4$, and $5$, the generating set of the triple product tensor is the same, with the given average fluctuations among the solutions, and so is the permutation symmetry group. See also the distribution of the generating set values in Figure \ref{fig:genSets_all}.

\begin{table}[h]
\centering\small
\begin{tabular}{@{}
  >{\centering\arraybackslash}p{3.2cm}
  >{\centering\arraybackslash}p{3.8cm}
  >{\centering\arraybackslash}p{3.2cm}
  >{\centering\arraybackslash}p{4.3cm}@{}}
\hline
\rule{0pt}{3.0ex}\textbf{Dimension} & $d = 3$ & $d = 4$ & $d = 5$ \\[2pt]
\hline
\rule{0pt}{3.0ex}\textbf{Generating Set}\newline $\mathrm{gen}(T_{ijk})$
  & $\{0,\ 0.5235987756,$ \newline $1.570796327,$ \newline $4.712388980, 5.759586532\}$ \newline
  & $\{0,\ 1.570796327,$ \newline $4.712388980\}$
  & $\{0,\ 0.6283185307,$ \newline $1.256637061,\ 3.141592654,$ \newline $5.026548246,\ 5.654866776\}$ \\[6pt]
\textbf{Avg. Fluctuation}
  & $\approx 10^{-15}$
  & $0$
  & $\approx 10^{-14}$ \newline \\[6pt]
\textbf{Symbolic Form}
  & $\left\{0,\,\tfrac{\pi}{6},\,\tfrac{\pi}{2},\,\tfrac{3\pi}{2},\,\tfrac{11\pi}{6}\right\}$
  & $\left\{0,\,\tfrac{\pi}{2},\,\tfrac{3\pi}{2}\right\}$
  & $\left\{0,\,\tfrac{\pi}{5},\,\tfrac{2\pi}{5},\,\pi,\,\tfrac{8\pi}{5},\,\tfrac{9\pi}{5}\right\}$ \newline \\[6pt]
\textbf{Frequency of Occurrence}
  & $\{1080,\,216,\,108,\,108,\,216\}$
  & $\{6080,\,960,\,960\}$
  & $\{13500,\,3000,\,3000,$ \newline $1500,\,3000,\,3000\}$ \newline \\[6pt]
$|\text{Aut}(T_{ijk})|$
  & $216$
  & $1920$
  & $3000$ \newline \\[6pt]
$\text{Aut}(T_{ijk})$
  & $\cong SL(2,\mathbb{Z}_{3}) \rtimes \mathbb{Z}_3^{2}$
  & $\cong \mathbb{Z}_2^4 \rtimes S_5$
  & $\cong SL(2,\mathbb{Z}_{5}) \rtimes \mathbb{Z}_5^{2}$ \\[2pt]
\hline
\end{tabular}
\caption{Characterization of generalized numerical MUB solutions found in dimensions $3$, $4$, and $5$. Generating set values are given up to $10$ digits; symbolic forms are identified within a tolerance of $10^{-13}$. Frequency of occurrence list the number of occurrences of each generating set element in the triple product tensor, in the same order as the generating set.}
\label{tab:characterization}
\end{table}

The automorphism groups of the generalized numerical solutions are the same as those of the standard analytical solution we investigated in Section \ref{section:symmetryKnownSols}. We verified this by utilizing the GAP software to infer the group structure of a few randomly chosen solutions in each dimension.

For all the solutions we have found in dimensions $3$, $4$ and $5$, we have performed isomorphism tests to check whether or not they are isomorphic to each other, in their respective dimensions. We have found that all the solutions are isomorphic. In other words, there exist permutations that map one triple product tensor of a solution to another. Therefore, no new form of inequivalent solution has been found in our search — verifying the claim of equivalence of all solutions in dimensions less than or equal to $5$ in \cite{Blanchfield2014}.

Moreover, following the method of Bruzda et al. \cite{Bruzda2017}, we have analyzed whether or not the solutions we have found are part of a continuous family of solutions on a smooth manifold of MUBs in the phase space constrained by the trace functions. We have found that all the solutions are isolated points in the phase space, with no continuous family of solutions existing around them.

\subsubsection*{\textbf{Dimension 6:}} The search in dimension $6$ was limited due to the significantly longer computation time. Table \ref{tab:constructionD6} summarizes the search results. Our search for a complete set of $d+1 = 7$ MUBs yields no valid solution: the constraint function stalls at a floor of $F(\Phi) \approx 10^{-2}$ across all trials, in contrast to the relatively rapid convergence to $F(\Phi) \approx 10^{-28}$ observed in dimensions $3$, $4$, and $5$ under identical settings.


\begin{table}[h]
\centering
\begin{tabular}{@{}cccccc@{}}
\hline
$d,m$ & \textbf{No.\ of Trials} & \textbf{Precision (digits)} & $F(\Phi)$ & \textbf{Avg.\ Time (s)} & \textbf{Success Rate} \\
\hline
$6,7$ & $10,000$ & $30$ & Min. $\approx 10^{-2}$ & $\approx 130$ & $0\%$ \\
$6,3$ & $300,000$ & $30$ & $\approx 10^{-28}$ & $\approx 0.4$ & $62\%$ \\
$6,4$ & $300,000$ & $30$ & Min. $\approx 10^{-2}$ & $\approx 1$  & $0\%$ \\
\hline
\end{tabular}
\caption{Summary of complete and incomplete MUB Gram matrix search results in dimensions $6$ with number of bases $3$ and $4$ and $7$. Out of $300,000$ trials $185,641$ valid solutions have been achieved in the $m=3$ case.}
\label{tab:constructionD6}
\end{table}

\begin{table}[h]
\centering\small
\renewcommand{\arraystretch}{1.5} 
\begin{tabular}{ c c c } 
\hline
\textbf{3 MUBs in Dimension 6} & & \\ 
\hline
$|\mathrm{gen}(T_{ijk})|$ & 9 & 73 \\
$|\mathrm{Aut}(T_{ijk})|$ & 72 & 3 \\
$\mathrm{Aut}(T_{ijk})$ & $\cong \mathbb{Z}_2^2 \times (\mathbb{Z}_3^2 \rtimes \mathbb{Z}_2)$ & $\cong \mathbb{Z}_3$ \\
\hline
\end{tabular}
\caption{Characterization of triple of MUB structures in dimension 6.}
\label{tab:characterizationD6}
\end{table}

The search result for incomplete number of MUBs is significant. Out of $300,000$ trials for 4 MUBs in dimension 6 — the existence of which remains a conundrum — no valid solutions that met the validity threshold have been found; in fact, the constraint function stalled around the same floor as that of search for complete set: $\approx 10^{-2}$.

On the other hand, triple MUBs solutions have a wider landscape. Out of $300,000$ trials, we have found $185,641$ MUB solutions with two distinct symmetry groups: about $8\%$ of solutions have a symmetry group of order $72$ isomorphic to $\mathbb{Z}_2^2 \times (\mathbb{Z}_3^2 \rtimes \mathbb{Z}_2)$ and about $92\%$ have symmetry group of order $3$ isomorphic to $\mathbb{Z}_3$. The latter MUB structure with symmetry group of order $3$ is rich in its variety of generating sets. Moreover, $30$-digit precision and $F \approx 10^{-28}$ were not enough to resolve the generating set of different solutions: the generating set elements were splitting up, artificially inflating the apparent size of the generating sets. Therefore, we refined the solutions up to $60$ digit precision and, consequently, down to the constraint value $F \approx 10^{-58}$. The size of the generating sets of all the solutions associated with $|\text{Aut}(T_{ijk})| = 3$ turned out to be $73$, though not necessarily consisting of the exact same elements. In contrast, the solutions associated with $|\text{Aut}(T_{ijk})| = 72$ are characterized by a single generating set of size $9$:
\begin{equation*}
\begin{aligned}
\{& 0, 0.2617993878, 0.7853981634, 1.3089969390,  2.3561944902,\\
&3.9269908170, 4.9741883682, 5.4977871438, 6.0213859194 \},
\end{aligned}
\end{equation*}
written here up to $10$ digits.

All the solutions characterized with $|\text{Aut}(T_{ijk})| = 72$ and $|\text{gen}(T_{ijk})| = 9$ are found to be isomorphic to each other, and are all isolated (i.e., $\Delta=0$). The rest of the solutions, associated with $|\text{Aut}(T_{ijk})| = 3$ and $|\text{gen}(T_{ijk})| = 73$, on the other hand, consists of inequivalent classes of MUBs and they are not isolated (with $\Delta=2$): perturbations around these solutions land to inequivalent solutions with similar automorphism group structure on the solution manifold.

For comparison with our generalized numerical constructions, we classify and investigate the neighbourhoods of several established analytic MUB triples from the literature. Besides the construction via the standard approach we reviewed in Section \ref{section:knownSols} which is also discussed in \cite{Grassl2009} specifically for the triples in dimension 6; we considered the explicit constructions in \cite{Goyeneche2013,McNulty2012}. The classification results are given in Table \ref{tab:characterizationKnownTriples}.

\begin{table}[h]
\centering\small
\renewcommand{\arraystretch}{1.5}
\begin{tabular}{@{}
  >{\centering\arraybackslash}p{3.2cm}
  >{\centering\arraybackslash}p{3.8cm}
  >{\centering\arraybackslash}p{3.2cm}
  >{\centering\arraybackslash}p{4.3cm}@{}}
\hline
\rule{0pt}{3.0ex}\textbf{Construction } & Standard WH & GEWH \cite{Goyeneche2013} & GMW \cite{McNulty2012} \\[2pt]
\hline
\rule{0pt}{3.0ex}$|\mathrm{gen}(T_{ijk})|$ & 9 & 13 & 9 \\
$|\mathrm{Aut}(T_{ijk})|$ & 72 & 24 & 36 \\
$\mathrm{Aut}(T_{ijk})$ & $\cong \mathbb{Z}_2^2 \times (\mathbb{Z}_3^2 \rtimes \mathbb{Z}_2)$ & $\cong (\mathbb{Z}_6 \times \mathbb{Z}_2) \rtimes \mathbb{Z}_2$ & $\cong \mathbb{Z}_2 \times (\mathbb{Z}_3^2 \rtimes \mathbb{Z}_2)$ \\
$\Delta$ & 0 & 2 & 4 \\[2pt]
\hline
\end{tabular}
\caption{Classification of various known triples of MUBs. GEWH refers to one form of Goyeneche's construction \cite{Goyeneche2013} and GMW is that of McNulty and Weigert's \cite{McNulty2012}.}
\label{tab:characterizationKnownTriples}
\end{table}

The two non-standard solutions shown in the Table \ref{tab:characterizationKnownTriples} have distinct group theoretical character. Both have nonzero restricted defect, which permits tangential directions along which further solutions may lie; we test this by first-order perturbation and refinement of each perturbed point back to a precision of $F \approx 10^{-50}$ via optimization with quasi-Newton method (see Section \ref{section:vicinity}). For GMW, whose automorphism group has order $36$ and whose restricted defect is $4$, only three of the four candidate directions are \textit{genuine}. Perturbation and refinement along the fourth direction does not reach any other solution but the original one. Therefore, the remaining three directions span a three-dimensional tangent space.

\begin{figure}[htbp]
    \centering
    \begin{subfigure}[b]{0.53\textwidth}
        \centering
        \includegraphics[width=\textwidth]{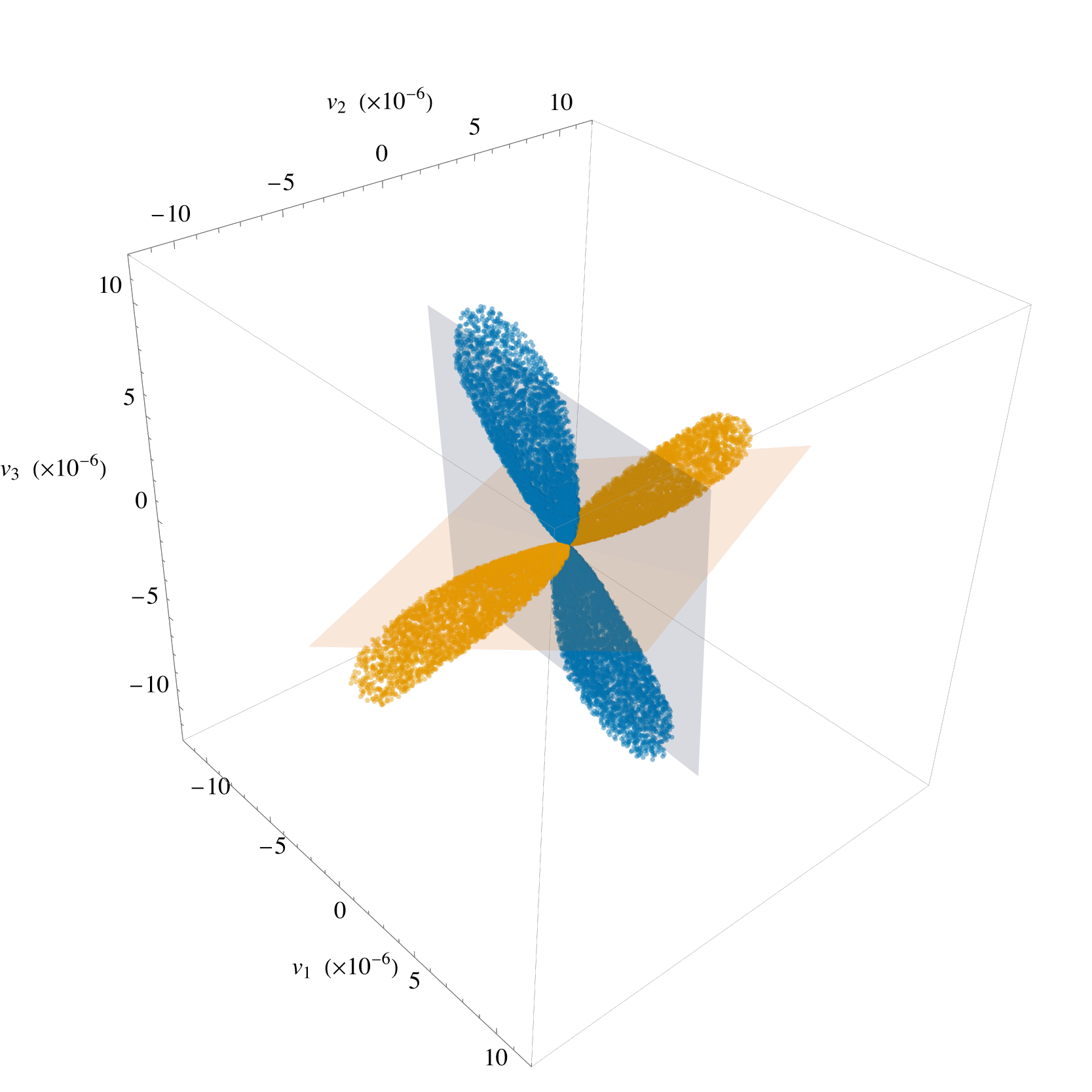}
        \label{fig:vicinity1a}
    \end{subfigure}
    \hfill
    \begin{subfigure}[b]{0.46\textwidth}
        \centering
        \includegraphics[width=\textwidth]{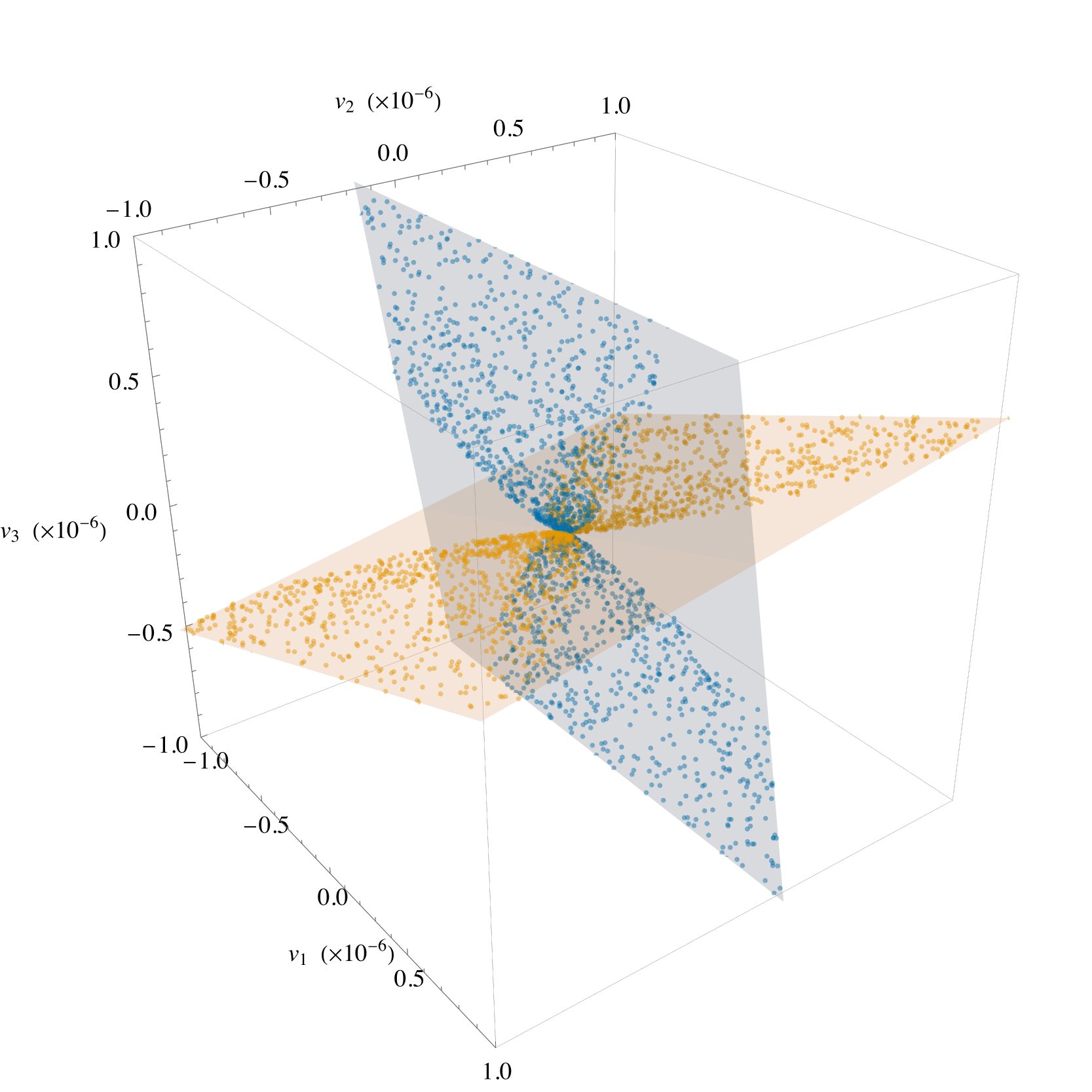}        \label{fig:vicinity1b}
    \end{subfigure}
    \caption{Solutions near GMW (origin) achieved by first-order perturbations with a step size of $10^{-5}$ along random directions in the tangent space and projected onto the tangent space at GMW, with $v_1$, $v_2$, $v_3$ the tangential basis directions. Plot ranges: (a) $\pm 10^{-5}$ (b) $10^{-6}$.}
    \label{fig:vicinity}
\end{figure}

Figure \ref{fig:vicinity} shows the resulting cluster of solutions, obtained from $20,000$ perturbations of GMW along directions drawn at random from this tangent space. Every perturbed solution is refined back to a $60$-precision digit such that $F \approx 10^{-50}$. The random first-order perturbations were taken with a step-size of $10^{-5}$. Each solution is a phase vector in the $108$-dimensional phase space, projected here onto the three nontrivial tangential directions. Every solution in the cluster has an automorphism group of order $3$. GMW, which our numerical search never produced, therefore sits as a rare solution at the intersection of two clusters of these far more abundant but less symmetric solutions.

The GEWH solution, whose automorphism group is of order $24$, behaves differently. Its restricted defect is $2$ and both candidate directions are genuine: first-order perturbations along them land on solutions. Unlike GMW, GEWH lies on a simple and smooth two-dimensional sheet of solutions rather than at an intersection of two planes. However, what the two share is that the surrounding solutions again all have automorphism groups of order $3$. The same behaviour holds for the solutions obtained from our generalized numerical construction with $\Delta = 2$: in every case we examined, the vicinity is a two-dimensional sheet.






\section{Conclusion}

In this work, we have developed a method of constructing generalized numerical MUBs based on the Hermitian and projection properties of the Gram matrix, expressed through the trace constraints. As the MUB Gram matrix constitutes a projection matrix, we have used the third and fourth order trace constraints to construct the geometry corresponding to MUBs. This method does not rely on any \textit{a priori} assumption of algebraic structure, such as finite fields or WH group. As such, generalized numerical construction allows us to explore the landscape of MUB solutions without any structural presumption that would restrict the MUB geometry to the known solutions.

Our framework for investigation and classification of MUBs through their geometric structure and symmetry properties — encoded in the triple products, generating set and the automorphism group of the triple product tensor — enables a classification of MUB solutions that goes beyond conventional unitary-equivalence arguments. The generating set, a set of unique triple products, captures the gauge-invariant signature and basis-independent geometry of the MUBs. This quality of a particular solution is a first step in identification of MUB structures, and distinguishing inequivalent MUBs. The automorphism group of the triple product tensor classifies the internal algebraic structure underlying a particular MUB solution.

Across all the constructions we have built, all complete MUB solutions in dimensions $3$, $4$, and $5$ are of the equivalent form and their symmetry groups are isomorphic to the normalizer of the WH group, i.e., the Clifford group; supporting the claim for the equivalence of complete MUBs up to dimension $5$ and their symmetry group being restricted to the Clifford group. Whereas a search that presupposes an algebraic structure can only recover what it presupposes, the generalized method provides an assumption-free analysis.

Our search for complete sets of MUBs in dimension $6$ is limited by computation times substantially longer than in lower dimensions, and within the range of our search we found no solution. However, configurations with incomplete set of MUBs are computationally faster. Our search for triple of MUBs reached several inequivalent solutions, whereas the search for four MUBs yielded no solutions. The existence of $4$ MUBs in dimension $6$ remains a long-standing open problem, and this, while not constituting a proof, reinforces the prevailing suspicion that $3$ is the maximal number of MUBs in dimension $6$. The persistence of the value $F(\Phi) \approx 10^{-2}$ under varying initialization for both complete set of $d+1=7$ MUBs and $4$ MUBs suggests that it is a feature of the geometry of the constraint landscape.

The two known triples we choose to classify appear to be rare solution points in the manifold which our numerical search would not produce in the given scope of exploration. Our analysis shows that the solution with automorphism group of order $36$ sits at the intersection of two planes of solution clusters formed by the more abundant form of solutions characterized with the automorphism group of order $3$. The solution with automorphism group of order $24$, although lies on a two-dimensional sheet of solutions, is populated with the solutions characterized with the similar automorphism group of order $3$ in its vicinity.




%
%

\section*{Acknowledgments}

Part of this work was carried out in Bilimler Köyü, Foça. The numerical computations were performed on Sabanci University High Performance Computing Sakura Cluster.

We would like to thank the editor and reviewers for their constructive feedbacks that made this work more complete.





\bibliographystyle{iopart-num}
\bibliography{library}

\end{document}